\documentclass{llncs}
\usepackage{graphicx}
\usepackage{color}
\usepackage{paralist}
\newcommand\NNG{\textsc{NeuGuS}}
\newcommand\sig{\mathtt{Ar}}
\newcommand\hoice{\textsc{HoIce}}
\newcommand\rtype{\texttt{r\_type}}
\newcommand\set[1]{\{#1\}}
\newcommand\Nat{\mathbf{Nat}}
\newcommand\Z{\mathbf{Z}}
\newcommand\form{\varphi}
\newcommand\sigmoid{\sigma}
\newcommand\imp{\Rightarrow}
\newcommand\Inv{\mathit{Inv}}
\newcommand\PV{\mathit{PVar}}
\newcommand\seq[1]{\widetilde{#1}}
\newcommand\Atoms[1]{\mathbf{Atoms}_{#1}}
\newcommand\mod{\mathbin{\mathit{mod}}}
\newcommand\N{\mathcal{N}}
\newcommand\loss[1]{\mathit{loss}_{#1}}
\newcommand\false{\mathtt{false}}
\newcommand\true{\mathtt{true}}
\newcommand\comment[1]{}
\newcommand\oracle{\mathtt{oracle}}

\newif\ifdraft\draftfalse
\newif\iffull\fulltrue
\ifdraft
\newcommand\nk[1]{\textcolor{red}{[#1 -nk]}}
\newcommand\TS[1]{\textcolor{cyan}{[#1 -ts]}}
\else
\newcommand\nk[1]{}
\newcommand\TS[1]{}
\fi

\makeatletter
\RequirePackage[bookmarks,unicode,colorlinks=true]{hyperref}%
   \def\@citecolor{blue}%
   \def\@urlcolor{blue}%
   \def\@linkcolor{blue}%

\def\orcidID#1{\smash{\href{http://orcid.org/#1}{\protect\raisebox{-1.25pt}{\protect\includegraphics{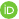}}}}}
\makeatother

\begin{document}
\title{Toward Neural-Network-Guided\\ Program Synthesis and Verification
\thanks{An Extended Version of the Summary in the Proceedings of SAS 2021, Springer LNCS.}}
%
%
\author{
  Naoki Kobayashi\inst{1}\orcidID{0000-0002-0537-0604}
  \and
  Taro Sekiyama\inst{2}\orcidID{0000-0001-9286-230X}
  \and
  Issei Sato\inst{1}
  \and
  Hiroshi Unno\inst{3,4}\orcidID{0000-0002-4225-8195}
  }

\authorrunning{N. Kobayashi et al.}
%
%
\institute{
  The University of Tokyo, Tokyo, Japan, \email{\{koba,issei.sato\}@is.s.u-tokyo.ac.jp}
  \and
  National Institute of Informatics \& SOKENDAI, Tokyo, Japan, \email{tsekiyama@acm.org}
  \and
  University of Tsukuba, Ibaraki, Japan
  \and
  RIKEN AIP, Tokyo, Japan
}

\maketitle              
\begin{abstract}
  We propose a novel framework of program and invariant synthesis
  called neural network-guided synthesis.
 We first show that, by suitably designing and training neural networks, 
  we can extract logical formulas over integers from
  the weights and biases of the trained neural networks.
Based on the idea, we have implemented a tool to synthesize formulas from positive/negative examples
and implication constraints, and obtained promising experimental results.
\iffull
  We also discuss two applications of our synthesis method.
  One is the use of our tool for qualifier discovery in the framework
  of ICE-learning-based CHC solving, which can in turn be applied to program verification
  and inductive invariant synthesis. Another application is to a new program development
  framework called oracle-based programming, which is a neural-network-guided
  variation of Solar-Lezama's program synthesis by sketching.
  \fi
\end{abstract}

%
%

\section{Introduction}
\label{sec:intro}
With the recent advance of machine learning techniques, there have been a lot of interests
in applying them to program synthesis and verification.
Garg et al.~\cite{garg_2014} have proposed the ICE-framework,
where the classical supervised learning based on positive and negative examples
been extended to deal with ``implication
constraints'' to infer inductive invariants.
Zhu et al.~\cite{DBLP:conf/pldi/ZhuXMJ19} proposed a novel approach to combining
neural networks (NNs)
and traditional software, where a NN controller is synthesized first, and
then an ordinary program that imitates the NN's behavior; the latter is used as a \emph{shield} for
the neural net controller and the shield (instead of the NN)
is verified by using traditional program verification techniques.
There have also been various approaches to directly verifying NN
components~\cite{DBLP:conf/sp/GehrMDTCV18,zhao2020learning,DBLP:conf/nips/AndersonVDC20,DBLP:conf/cav/PulinaT10,DBLP:conf/aaai/NarodytskaKRSW18}.

We propose yet another approach to using neural networks for program
verification and synthesis.  Unlike the previous approaches where
neural networks are used either as black
boxes~\cite{DBLP:conf/pldi/ZhuXMJ19,DBLP:conf/icml/VermaMSKC18} or
white boxes~\cite{DBLP:conf/sp/GehrMDTCV18,DBLP:conf/cav/KatzBDJK17},
our approach treats neural networks as \emph{gray} boxes.  Given training
data,
which typically consist of input/output examples for
a (quantifier-free) logical formula (as a part of a program
component or a program invariant) to be synthesized, we first train a NN.
We then synthesize a logical formula
by using the weights and biases of the trained NN as hints.
Extracting simple (or, ``interpretable''), classical\footnote{Since neural networks can also be expressed as programs,
  we call ordinary programs written without using
  neural networks \emph{classical}, to distinguish them from
  programs containing NNs.} 
program expressions from NNs
has been considered difficult, especially for
deep NNs; 
in fact, achieving ``explainable AI''~\cite{DBLP:journals/inffus/ArrietaRSBTBGGM20}
has been a grand challenge in the field of machine learning. 
Our thesis here is, however, that if NNs are suitably designed with program or invariant synthesis in mind,
and if the domain of the synthesis problems is suitably restricted to those
which have reasonably simple program expressions as solutions,
then it is actually often possible to extract program expressions (or logical formulas)
by inspecting the weights of trained NNs.

\begin{figure}[tbp]
  \begin{center}
    \includegraphics[scale=0.4]{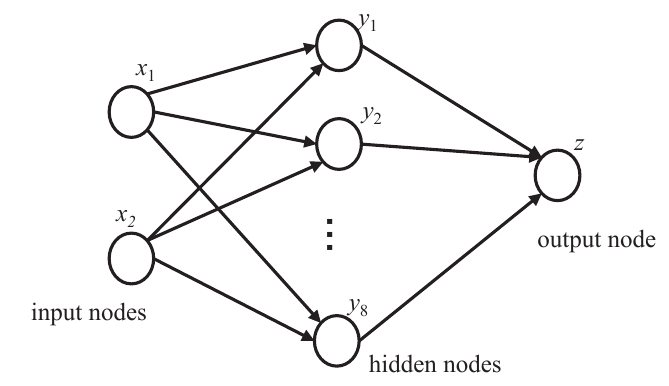}
    \vspace*{-.5cm}
  \end{center}
  \caption{A Neural Network with One Hidden Layer}
  \label{fig:nn}
\end{figure}

To clarify our approach, we give an example of the extraction.
Let us consider the three-layer neural network shown in Figure~\ref{fig:nn}.
The NN is supposed to work as a binary classifier for two-dimensional
data: it takes a pair of numbers \((x_1,x_2)\) as an input,
and outputs a single number \(z\), which is expected to be a value close to either
\(1\) or \(0\). The NN has eight hidden nodes, and the sigmoid function \(\sigmoid\)
is used as activation functions for both the hidden and output
nodes.\footnote{These design details can affect the efficacy of our program
  expression extraction, as discussed later.}
The lefthand side of Table~\ref{fig:training-example} shows a training data set,
where each row consists of inputs 
\((x_1,x_2)\ (-25\le x_1, x_2\le 25)\), and their 
labels, which are \(1\) if \(4x_1+x_2>0\land 2x_1+3 x_2+9<0\).
The righthand side of Table~\ref{fig:training-example} shows
the weights and biases of the trained NN.
The \(i\)-th row
shows information about each hidden node: \(w_{1,i}\), \(w_{2,i}\), and \(b_i\)
(\(i\in\set{1,\ldots,8}\))
are the weights and the bias for
 the links connecting the two input nodes and the \(i\)-th hidden node, and
\(w_{o,i}\) is the weight for the link connecting the hidden node and the output node.
(Thus, the value \(y_i\) of the \(i\)-th hidden node for inputs \((x_1,x_2)\) is
\(y_i=\sigmoid(b_i+w_{1,i}x_1+w_{2,i}x_2)\), and the 
output of the whole network is
\(\sigmoid(b_o+\sum_{i=1}^{8}w_{o,i}y_i)\) for some bias $b_o$.)
The rows are sorted according to the absolute value of \(w_{o,i}\).
We can observe that the ratios among \(w_1, w_2, b\) of the first four rows are roughly
\(2:3:9\), and those of the last four rows are roughly \(4:1:0\).
Thus, the value of each hidden node is close to \(\sigmoid(\beta(2x_1+3x_2+9))\)
or \(\sigmoid(\beta(4x_1+x_2))\) for some \(\beta\). Due 
to the property that the value of the sigmoid function \(\sigmoid(u)\)
is close to \(0\) or \(1\) except near \(u=0\), 
we can guess that \(2x_1+3x_2+9 \diamond 0\) and
\(4x_1+x_2 \diamond 0\) (where \(\diamond \in \set{<, >, \le, \ge}\))
are relevant to the classification. Once the relevant atomic formulas are obtained,
we can easily find the correct classifier \(4x_1+x_2>0\land 2x_1+3 x_2+9<0\) by solving
the problem of Boolean function synthesis in a classical manner.

\begin{table}[tbp]
  \caption{Training Data (left) and the Result of Learning (right)}
  \label{fig:training-example}
  \begin{tabular}{|rrr|}
    \hline
$x_1$ &\hspace*{1em}     $x_2$ &\hspace*{.5em}    label\\
  \hline
 $ 25 $ & $ 25 $ & $ 0$ \\
 \(\cdots\) &  &  \\
 $ 25 $ & $ -21 $ & $ 0$ \\
 $ 25 $ & $ -20 $ & $ 1$ \\
 $25 $ & $ -21 $ & $ 1$ \\
 \(\cdots\) &  &  \\
 $25 $ & $ -25$ & $ 1$ \\
 $24$ & $ 25$ & $ 0$ \\
 \(\cdots\) &  &  \\
 $-25$ & $ -25$ & $ 0$ \\
    \hline
  \end{tabular}
  \begin{tabular}{|rrrr|}
    \hline
    \(w_{1,i}\) & \(w_{2,i}\) & \(b_i\) & \(w_{o,i}\)\\
    \hline
    \hline
\(4.037725448 \) &\quad \( 6.056035518 \) &\quad \(  18.18252372 \) &\quad \( -11.76355457\)\\
\(4.185569763 \) & \( 6.27788019 \) & \( 18.92045211 \) & \( -11.36994552\)\\
\(3.775603055 \) & \( 5.662680149 \) & \( 16.86475944 \) & \( -10.83486366\)\\
\(3.928676843 \) & \( 5.892404079 \) & \( 17.63601112 \) & \( -10.78136634\)\\
\(-15.02299022 \) & \(  -3.758415699 \) & \(  1.357473373 \) & \(    -9.199707984\)\\
\(-13.6469354 \) & \(  -3.414942979 \) & \(  1.145593643 \) & \(    -8.159229278\)\\
\(-11.69845199 \) & \(  -2.927870512 \) & \(  0.8412334322 \) & \(   -7.779587745\)\\
\(-12.65479946 \) & \(  -3.168056249 \) & \(  0.9739738106 \) & \(   -6.938682556\)\\
\hline
\end{tabular}
\end{table}

We envision two kinds of applications of our NN-guided predicate synthesis
sketched above.
One is to an ICE-based learning of inductive program invariants~\cite{garg_2014,DBLP:journals/jar/ChampionCKS20}. One of the main bottlenecks of the ICE-based learning method (as in 
many other 
verification methods) has been the discovery of appropriate qualifiers (i.e. atomic
predicates that constitute invariants). Our NN-guided predicate synthesis can be used
to find qualifiers, by which reducing the bottleneck.
The other potential application 
is to a new program development framework called
\emph{oracle-based programming}. It is a neural-network-guided
variation of Solar-Lezama's program synthesis by sketching~\cite{solar2008program}.
As in the program sketch,
a user gives a ``sketch''
of a program to be synthesized, and 
a 
synthesizer tries to find expressions to fill the holes
(or \emph{oracles} in our terminology) in the sketch.
By using our method outlined above, we can first prepare a NN for each hole,
and train the NN by using data collected from the program sketch and specification.
We can then guess an expression for the hole from the weights of the trained NN.

The contributions of this paper are:
(i) the NN-guided predicate synthesis method sketched above,
(ii) experiments that confirm the effectiveness of the proposed method,
and (iii) discussion and preliminary experiments
on the potential applications to program verification and synthesis mentioned above.

Our idea of extracting useful information from NNs resembles that of symbolic
regression and extrapolation~\cite{DBLP:conf/iclr/MartiusL17,Petersen21ICLR,pmlr-v80-sahoo18a},
where domain-specific networks are designed and trained to learn
mathematical expressions. Ryan et al.~\cite{DBLP:conf/iclr/RyanWYGJ20,DBLP:conf/pldi/YaoRWJG20}
recently proposed logical regression to learn SMT formulas and applied it to the discovery of loop invariants.
The main differences from those studies are: (i) our approach is hybrid: we use NNs as gray boxes
to learn relevant inequalities, and combine it with classical methods for Boolean function learning, and
(ii) our learning framework is more general in that it takes into account positive/negative samples, and implication
constraints; see Section~\ref{sec:rel} for more discussion.

The rest of this paper is structured as follows.
Section~\ref{sec:PN} shows our basic method for synthesizing logical
formulas from positive and negative examples, and reports experimental results.
Section~\ref{sec:ICE} extends the method to deal with
\emph{implication constraints} in the ICE-learning framework~\cite{garg_2014},
and discusses an application to CHC (Constrained Horn Clauses) solving.
\iffull
Section~\ref{sec:oracle} introduces the new framework of oracle-based programming,
and shows how our synthesis method can be used as a key building block.
\fi
Section~\ref{sec:rel} discusses related work and Section~\ref{sec:conc} concludes
the paper.


\section{Predicate Synthesis from Positive/Negative Examples}
\label{sec:PN}

In this section, we propose a method for synthesizing logical formulas on integers
from positive and negative examples, and report experimental results.
 We will extend the method
 to deal with ``implication constraints''~\cite{garg_2014,DBLP:journals/jar/ChampionCKS20}
 in Section~\ref{sec:ICE}.

 \subsection{The Problem Definition and Our Method}
 \label{sec:method}
The goal of our synthesis is defined as follows. 
\begin{definition}[Predicate synthesis problem with P/N Examples]
  The predicate synthesis problem with positive (P) and negative (N) examples (the PN synthesis problem,
  for short) is, given sets \(P, N\subseteq \Z^k\) of positive and negative examples
  (where \(\Z\) is the set of integers) such that \(P\cap N=\emptyset\),
  to find a logical formula \(\form(x_1,\ldots,x_k)\) such that
  \(\models \form(v_1,\ldots,v_k)\) holds for every \((v_1,\ldots,v_k)\in P\)
  and 
  \(\models \neg\form(v_1,\ldots,v_k)\) for every \((v_1,\ldots,v_k)\in N\).
\end{definition}
\TS{We could briefly explain positive and negative examples can be given in considered applications (CHC solving and oracle-based programming).}\nk{I don't think that is necessary.}
For the moment, we assume that formulas are those of linear integer
arithmetic (i.e. arbitrary Boolean combinations of linear integer inequalities);
some extensions to deal with the modulo operations and
polynomial constraints will be discussed later in Section~\ref{sec:nonlinear}.

\begin{figure}[tbp]
  \begin{center}
    \includegraphics[scale=0.5]{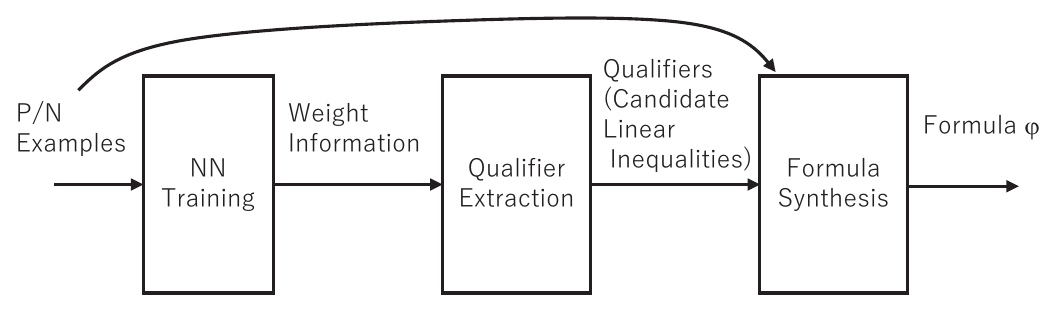}
  \end{center}
  \caption{An Overall Flow of Our Synthesis Framework}
  \label{fig:flow}
\end{figure}

The overall flow of our method is depicted in Figure~\ref{fig:flow}.
We first (in the NN training phase) train a NN on a given set of P/N examples,
and then (in the qualifier extraction
phase) extract linear inequality constraints (which we call ``qualifiers'')
by inspecting 
the weights and biases of the trained NN, as sketched in Section~\ref{sec:intro}.
Finally (in the formula synthesis phase), we construct a Boolean combination
of the qualifiers that matches P/N examples. Note that the trained NN is
used only for qualifier extraction; in the last phase, we
use a classical algorithm
for Boolean function learning.

\begin{figure}[tbp]
  \begin{center}
    \includegraphics[scale=0.4]{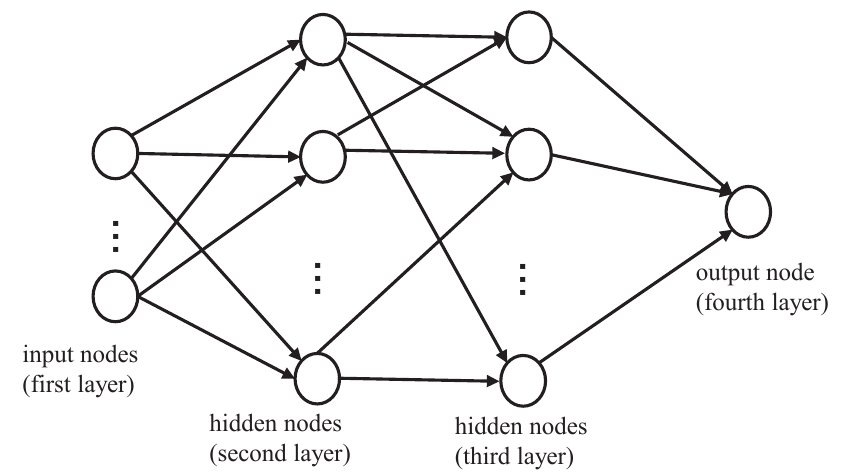}
  \end{center}
 \vspace*{-3ex}
  \caption{A Four-Layer Neural Network}
  \label{fig:3layerNN}
\end{figure}

Each phase is described in more detail below.
\paragraph{NN Training.}
We advocate the use of a four-layer neural network
as depicted in Figure~\ref{fig:3layerNN},
where the sigmoid function \(\sigmoid(x) = \displaystyle\frac{1}{1+e^{-x}}\)
is used as the activation functions for all the nodes.

We briefly explain the four-layer neural network;
those who are familiar with neural networks
may skip this paragraph.
Let us write \(N_{i,j}\) for the \(j\)-th node in the \(i\)-th layer,
and \(m_i\) for the number of the nodes in the \(i\)-th layer (hence, \(m_4\),
which is the number of output nodes, is \(1\)).
Each link connecting \(N_{i-1,j}\) and \(N_{i,k}\) (\(i\in \set{2,3,4}\))
has a real number called the \emph{weight} \(w_{i,j,k}\), and
each node \(N_{i,j}\) (\(i\in \set{2,3,4}\)) also has another real number
\(b_{i,j}\) called the \emph{bias}.
The value \(o_{i,j}\) of each node \(N_{i,j}\) (\(i>1\)) is calculated from
the input values \(o_{0,j}\) by the equation:
\[
o_{i,j} = f\big(b_{i,j}+\sum_{k=1}^{m_{i-1}} w_{i,k,j}\cdot o_{i-1,k}\big),
\]
where the function \(f\), called an activation function,
is the sigmoid function \(\sigmoid\) here; other popular activation functions include
\(\mathtt{ReLU}(x)=max(0, x)\) and \(\mathtt{tanh}=\displaystyle\frac{e^x-e^{-x}}{e^x+e^{-x}}\).
The weights and biases are updated during the training phase.
The training data are a set of pairs \((d_i,\ell_i)\)
where \(d_i=(o_{1,1},\ldots,o_{1,m_1})\) is an input and
\(\ell_i\) is a label (which is \(1\) for a positive example and \(0\) for
a negative one).
The goal of training a NN is to adjust the weights
and biases to (locally) minimize
the discrepancy between the output \(o_{4,1}\) of the NN and \(\ell_i\)
for each \(d_i\). That is usually achieved by defining an appropriate
\emph{loss} function for the training data, and repeatedly updating the
weights and biases in a direction to decrease the value of the loss function
by using the gradient descent method.

Our intention in choosing the four-layer NN in Figure~\ref{fig:3layerNN}
is to force the NN to recognize qualifiers
(i.e., linear inequality constraints) in the second layer (i.e., the first
hidden layer),
and to recognize an appropriate Boolean combination of them in the
third and fourth layers.
The sigmoid function was chosen as the activation function of the second
layer to make it difficult for the NN to propagate information
on the inputs besides information about linear inequalities,
so that we can extract linear inequalities only by looking at the weights and
biases for the hidden nodes in the second layer.
Note that the output of each hidden node in the second layer is of the form:
\[ \sigmoid(b+w_1\,x_1+\cdots+w_k\,x_k).\]
Since the output of the sigmoid function 
is very close to \(0\) or \(1\)
except around \(b+w_1\,x_1+\cdots+w_k\,x_k=0\)
and input data \(x_1,\ldots,x_k\) take discrete values,
 only information
about \(b+w_1\,x_1+\cdots+w_k\,x_k>c\) for small \(c\)'s may be propagated
to the second layer. In other words, the hidden nodes in the second layer
can be expected to recognize ``features'' of the form
\(b+w_1\,x_1+\cdots+w_k\,x_k>c\), just like the initial layers of DNNs for image
recognition tend to recognize basic features such as lines.
The third and fourth layers are intended to recognize a (positive) Boolean combination
of qualifiers. We expect the use of two layers (instead of only one layer) for this task
makes it easier for NN to recognize
Boolean formulas in DNF or CNF.\footnote{According to the experimental results
  reported later, however, the three-layer NN as depicted in Figure~\ref{fig:nn}
  also seems to be a good alternative. It is left for future work to test
  whether NNs with more than four layers are useful for our task.}
Notice that the conjunction \(x_1\land \cdots \land x_k\)
and the disjunction \(x_1\lor \cdots \lor x_k\)
can respectively be approximated by
\(\sigmoid(\beta(x_1+\cdots+x_k-\frac{2k-1}{2}))\) and
\(\sigmoid(\beta(x_1+\cdots+x_k-\frac{1}{2}))\)
for a sufficiently large positive number \(\beta\).

For the loss function of the NN, there are various candidates.
In the experiments reported later, we tested the mean square error function
(the average of \((o_i - \ell_i)^2\), where \(o_i\) is the output of
NN for the \(i\)-th training data, and \(\ell_i\) is the corresponding label)
and the mean of a logarithmic error function
(the average of \(-\log(1-|o_i-\ell_i|)\)).\footnote{We actually used
  \(-\log(\max(1-|o_i-\ell_i|,\epsilon))\) for a small positive number \(\epsilon\), to avoid
  the overflow of floating point arithmetic.}

\paragraph{Qualifier Extraction.}
From the trained NN, for each hidden node \(N_{2,i}\) in the second layer,
we extract the bias \(b_{2,i}\) (which we write
\(w_{2,0,i}\) below for technical convenience)
and weights \(w_{2,1,i},\ldots,w_{2,k,i}\), and
construct integer coefficients \(c_0,c_1,\ldots,c_k\) such that
the ratio \(c_n/c_m\) of each pair of coefficients is close to
\(w_{2,n,i}/w_{2,m,i}\).
We then generate linear inequalities of the form 
\(c_0+c_1x_1+\cdots+c_kx_k>e\) where \(e\in \set{-1,0,1}\) as qualifiers.
The problem of obtaining the integer coefficients \(c_0,\ldots,c_k\)
is essentially a Diophantine approximation problem,
which can be solved, for example, by using continued fraction expansion.
The current implementation uses the following naive, ad hoc method.\footnote{
We plan to replace this with a more standard one based on continued fraction
expansion.}
We assume that the coefficients of relevant qualifiers are integers smaller than
a certain value \(K\) (\(K=5\) in the experiments below).
Given \(w_0+w_1x_1+\cdots+w_kx_k\), we pick \(i>0\) such that
\(|w_i|\) is the largest among \(|w_1|,\ldots,|w_k|\), and normalize
it to the form \(w'_0 + w_1' x_1+\cdots + w_k' x_k\) where \(w_j'=w_j/w_i\)
(thus, \(0\le |w_j'|\le 1\) for \(j>0\)).
We then pick \(-K< n_j,m_j< K\) such that \(n_j/m_j\) is closest to \(|w_j'|\),
and obtain \(w'_0 + (n_1/m_1) x_1 + \cdots + (n_k/m_k) x_k\).
Finally, by multiplying the expression
with the least common multiple \(M\) of \(m_1,\ldots, m_k\),
and rounding \(M w'_0\) with an integer \(c_0\), we obtain
\(c_0 + M(n_1/m_1) x_1 + \cdots + M(n_k/m_k) x_k\).

If too many qualifiers are extracted (typically when
the number of hidden nodes is large;
note that qualifiers are extracted from each hidden node),
we prioritize them by inspecting the weights of the third and fourth layers,
and possibly discard those with low priorities. The priority
of qualifiers obtained from the \(i\)-th hidden node \(N_{2,i}\)
in the second layer is set
to \(p_{i}=\sum_{j} |w_{3,i,j}w_{4,j,1}|\),
The priority \(p_i\) estimates
 the influence of the value of the \(i\)-th hidden node,
hence the importance of the corresponding qualifiers.

\paragraph{Formula Synthesis.}
This phase no longer uses the trained NN and simply applies
a classical method (e.g. the Quine-McCluskey method, if
the number of candidate qualifiers is small)
for synthesizing a Boolean formula from a truth table
(with don't care inputs).\footnote{The current implementation uses an ad hoc,
  greedy method, which will be replaced by a more standard one for
  Boolean decision tree construction.}
Given qualifiers \(Q_1,\ldots,Q_m\) and P/N examples \(d_1,\ldots,d_n\) with
their labels \(\ell_1,\ldots,\ell_n\),
the goal is to construct a Boolean combination \(\form\) of \(Q_1,\ldots,Q_m\) such that
\(\form(d_i)=\ell_i\) for every \(i\). To this end,
we just need to construct a truth table where each row consists of
\(Q_1(d_i),\ldots,Q_m(d_i), \ell_i\) (\(i=1,\ldots,n\)),
and obtain a Boolean function \(f(b_1,\ldots,b_m)\)
such that \(f(Q_1(d_i),\ldots, Q_m(d_i))=\ell_i\) for every \(i\), and
let \(\form\) be \(f(Q_1,\ldots,Q_m)\). Table~\ref{tab:synthesis}
gives an example,
where \(m=2\), \(Q_1=x+2y>0\) and \(Q_2=x-y<0\). In this case, \(f(b_1,b_2)=b_1\land b_2\),
hence \(\form= x+2y>0\land x-y<0\).

One may think that we can also use information on the weights of the third and 
fourth layers of the trained NN in the formula synthesis phase.
Our rationale for not doing so is as follows.
\begin{asparaitem}
\item Classically (i.e. without using NNs) synthesizing a Boolean function is
  not so costly; it is relatively much cheaper than the task of finding relevant qualifiers
  in the previous phase.
\item It is not so easy to determine from the weights what Boolean formula is represented by
  each node of a NN. For example, as discussed earlier,
\(x\land y\) and \(x\lor y\) can be represented by
\(\sigmoid(\beta(x+y-\frac{3}{2}))\) and \(\sigmoid(\beta(x+y-\frac{1}{2}))\),
whose difference is subtle (only the additive constants differ).
\item By only using partial information about the trained NN, we need not worry too
  much about the problem of overfitting\TS{Why?}\nk{That is explained in ``By only ...''
    and ``Indeed, ...''}.
  Indeed, as discussed later in Section~\ref{sec:exp}
  and Appendix~\ref{sec:twolayer-patchwork}, a form of overfitting is observed also in our experiments,
  but even in such cases, we could extract useful qualifiers.
\item As confirmed by the experiments reported later,
  not all necessary qualifiers may be extracted in the previous phase;
  in such a case,
  trying to extract a Boolean formula directly from the NN would fail.
   By using a classical approach to a Boolean function synthesis,
   we can easily take into account the qualifiers collected by other means
  (e.g., in the context of program invariant synthesis, we often
  collect relevant qualifiers from
  conditional expressions in a given program).
\end{asparaitem}

Nevertheless,
it is reasonable to use the weights of the second and third layers
of the trained NN as \emph{hints} for the synthesis of \(\form\),
which is left for future work.

\begin{table}[tbp]
  \caption{An Example of Truth Tables Constructed from Qualifiers and Examples}
  \label{tab:synthesis}
\begin{center}
\begin{tabular}{|l|l|l|l|}
  \hline
  $d_i$ & $x+2y>0$ & $x-y<0$ & $\ell_i$\\
  \hline
  (1,0) & 1 & 1 & 1\\
  (1,1) & 1 & 0 & 0\\
  $\cdots$ &   $\cdots$ &   $\cdots$ &   $\cdots$ \\
  (-2,-1) & 0 & 1 & 0\\
  (-2,-2) & 0 & 0 & 0\\
  \hline
\end{tabular}
\end{center}
\end{table}

\subsection{Experiments}
\label{sec:exp}
We have implemented a tool called \NNG{}
based on 
the method described above using ocaml-torch\footnote{\url{https://github.com/LaurentMazare/ocaml-torch}.}, OCaml interface for the PyTorch library,
and conducted experiments. The source code of our tool is available at
\url{https://github.com/naokikob/neugus}.
All the experiments were conducted on a laptop computer with Intel(R) Core(TM) i7-8650U CPU (1.90GHz) and 16 GB memory; GPU was
not utilized for training NNs.
\subsubsection{Learning Conjunctive Formulas.}
\begin{figure}[tbp]
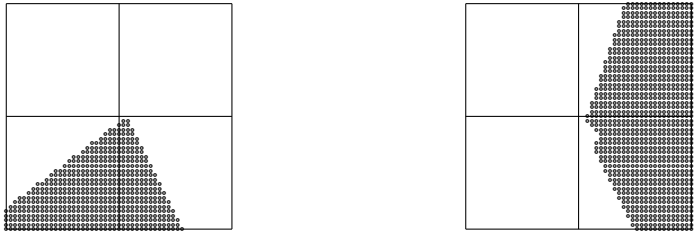

  \begin{center}
\begin{minipage}{6cm}
\unitlength=0.3mm

\end{minipage}
\end{center}
\caption{Visualization of sample instances: \(-2x-y+4>0\land 3x-4y+5>0\) (left) and
  $(2x+y+1>0\land x-y-9>0)\lor (x+y>0\land 3x-y-5>0)$ (right, Instance \#10 for
  \((A\land B)\lor (C\land D)\)).
  The small circles show positive examples, and the white area is filed with negative examples.}
  \label{fig:data2d}
\end{figure}

As a warming-up, we have randomly generated a conjunctive formula \(\form(x,y)\) of the form \(A\land B\)
where \(A\) and \(B\) are linear inequalities of the form \(a x+ b y + c>0\), and
\(a, b\) and \(c\) are integers such that
\(-4 \le a, b\le 4\) and \(-9\le c\le 9\) with \(ab\ne 0\).
We set \(P = \set{(x,y)\in \Z^2 \cap ([-25,25]\times[-25,25]) \mid \models \form(x,y)}\)
and \(N = \set{(x,y)\in \Z^2 \cap ([-25,25]\times[-25,25]) \mid \models \neg\form(x,y)}\)
as the sets of positive and negative examples, respectively
(thus, \(|P\cup N| = 51\times 51 = 2601\)).\footnote{We have excluded out
  instances where \(A\) or \(B\) is subsumed by the other, and those where the set
  of positive or negative examples is too small.}
The left-hand side of Figure~\ref{fig:data2d} plots positive examples for
\(A\equiv -2x-y+4>0\) and \(B\equiv 3x-4y+5>0\).
We have randomly generated 20 such formula instances, and ran our tool three times for each instance
to check whether the tool could find qualifiers \(A\) and \(B\).
In each run, the NN training was repeated either until the accuracy
becomes 100\% and the loss becomes small enough,
or until the number of training steps
(the number of forward and backward propagations) reaches 30,000.
If the accuracy does not reach 100\% within 30,000 steps or the loss does not become
small enough (less than \(10^{-4}\)),
the tool retries the NN training from scratch, up to three times.
(Even if the accuracy does not reach 100\% after three retries, the tool proceeds
to the next phase for qualifier discovery.)
As the optimizer, we have used Adam~\cite{DBLP:journals/corr/KingmaB14}
with the default setting of
ocaml-torch (\(\beta_1=0.9\), \(\beta_2=0.999\), no weight decay), and
the learning rate was \(0.001\). We did not use mini-batch training; all
the training data were given at each training step.
  
Table~\ref{tab:learn-conj} shows the result of experiments.
The meaning of each column is as follows:
\begin{itemize}
  \item ``\#hidden nodes'': the number of hidden nodes.
``$m_1\mathbin{:}m_2$'' means that a four-layer NN was used, and that the numbers of hidden nodes in the second
and third layers are respectively \(m_1\) and \(m_2\),
while ``$m$'' means that a three-layer (instead of four-layer) NN was used,
with the number of hidden nodes being \(m\).
\item ``loss func.'': the loss function used for training.
  ``log'' means  \(\displaystyle\frac{1}{n}\sum_{i=1}^n -\log(1-|o_i-\ell_i|)\)
  (where \(o_i\) and \(\ell_i\) are the prediction and label for the \(i\)-th example),
  and ``mse'' means the mean square error function.
\item ``\#retry'': the total number of retries. 
For each problem instance, up to 3 retries were performed.
  (Thus, there can be \(20 \mbox{ (instances) }\times 3 \mbox{ (runs) } \times 3 \mbox{ (retries per run) } = 180\)
  retires in total at maximum.)
\item ``\%success'': the percentage of runs in which a logical formula that
  separates positive and negative examples was constructed.
  The formula may not
  be identical to the original formula used to generate the P/N examples
  (though, for this experiment of
    synthesizing \(A\land B\), all the formulas synthesized were identical to the original formulas).
\item ``\%qualifiers'': the percentage
  of the original qualifiers (i.e.,
  inequalities \(A\) and \(B\) in this experiment) found.
\item ``\#candidates'': the average number of qualifier candidates extracted from
  the NN. Recall that from each hidden node in the first layer, we extract three inequalities
  (\(c_0+c_1x_1+\cdots +c_kx_k > e\) for \(e\in\set{0,-1,1}\)); thus, the maximum number of generated candidates
is \(3m\) for a NN with \(m\) hidden nodes in the second layer.
\item ``time'': the average execution time per run. Note that the current implementation is naive and does not use GPU. 
\end{itemize}
As shown in Table~\ref{tab:learn-conj}, our tool worked quite well;
it could always find the original
qualifiers, no matter whether the number of layers is three or four.\footnote{We have
  actually tested our tool also with a larger number of nodes, but we omit those results since
  they were the same as the case for 4:4 and 4 shown in the table: 100\% success rate and 100\% qualifiers found.}

\begin{table}[tbp]
\caption{Experimental Results for Learning a Conjunctive Formula \(A\land B\).}
\label{tab:learn-conj}
  \begin{center}
\begin{tabular}{|c|c|c|c|c|c|c|}
\hline   
\#hidden nodes & loss func. & \#retry & \%success & \%qualifiers & \#candidates & time (sec.)\\
\hline
\hline
4:4 & log & 0 & 100\% & 100\% & 6.8 & 27.7\\
\hline
4 & log & 0 & 100\% & 100\% & 6.7 & 25.2\\
\hline
\end{tabular}
\end{center}
\end{table}

\subsubsection{Learning Formulas with Conjunction and Disjunction}
We have also tested our tool for formulas with both conjunction and disjunction.
We have randomly
generated 20 formulas of the form \((A\land B)\lor (C\land D)\),\footnote{After the generation, we have manually excluded
  instances that have simpler equivalent formulas (e.g. \((x+y>1\land x+y>0)\lor (x-y>1\land x-y>0)\) is equivalent
to \(x+y>1\lor x-y>1\), hence removed), and regenerated formulas.}
where \(A,B,C\), and \(D\) are linear inequalities,
and prepared the sets of positive and negative examples
as in the previous experiment.
The right-hand side of Figure~\ref{fig:data2d} plots positive examples for
$(2x+y+1>0\land x-y-9>0)\lor (x+y>0\land 3x-y-5>0)$.
\iffull
The visualization of all the 20 instances is given in Appendix~\ref{sec:2d4instances}.
\fi
As before, we ran our tool three times for each instance, with several variations of NNs.

The result of the experiment is summarized in Table~\ref{tab:learn-dnf}, where the meaning of each column is the same
as that in the previous experiment. Among four-layer NNs, 32:32 with the log loss function
showed the best performance in terms of the columns \%success
and \%qualifiers. However, 8:8 with the log function also performed very well, and is preferable in terms of
the number of qualifier candidates generated (\#candidates). As for the two loss functions, the log function generally
performed better; therefore, we use the log function in the rest of the experiments. The running time does not vary
among the variations of NNs; the number of retries was the main factor to determine the running time.

\begin{table}[tbp]
\caption{Experimental Results for Learning Formulas of the form \((A\land B)\lor (C\land D)\).}
\label{tab:learn-dnf}
  \begin{center}
\begin{tabular}{|c|c|c|c|c|c|c|}
\hline   
\#hidden nodes & loss func. & \#retry & \%success & \%qualifiers & \#candidates
& time (sec.)\\
\hline
\hline
8:8 & log & 12 & 93.3\% & 96.7\% & 18.6 & 42.1\\
\hline
8:8 & mse & 17 & 86.7\% & 94.6\% & 18.1 & 40.6\\
\hline
16:16 & log & 0 & 91.7\% & 95.8\% & 26.2 & 32.9\\
\hline
16:16 & mse & 3 & 85.0\% & 92.5\% & 26.9 & 31.7\\
\hline
32:32 & log & 0 & 100\% & 97.9\% & 38.9  & 31.1\\
\hline
32:32 & mse & 15 & 85.0\% & 91.7\% & 40.4 & 39.9\\
\hline
8 & log & 156 & 36.7\% & 66.3\% & 21.6 & 92.4\\
\hline
16 & log & 54 & 95.0\% & 96.7\% & 37.1 & 55.8\\
\hline
32 & log & 2  & 100\% & 98.3\% & 58.5 & 36.3\\
\hline
\end{tabular}
\end{center}
\end{table}

As for three-layer NNs, the NN with 8 hidden nodes performed quite badly. This matches the intuition that
the third layer alone is not sufficient for
recognizing the nested conjunctions and disjunctions.
To our surprise, however, the NN with 32 hidden nodes actually showed the best performance in terms of
\%success and \%qualifiers, although \#candidates (smaller is better) is much larger than those for four-layer NNs.
We have inspected the reason for this, and found that the three-layer
NN is not recognizing
\((A\land B)\lor (C\land D)\), but classify positive and negative examples by doing a kind of patchwork, using a large number of (seemingly irrelevant) qualifiers which happened to include the correct qualifiers \(A,B,C\), and \(D\);
for interested readers, we report our analysis in Appendix~\ref{sec:twolayer-patchwork}.
We believe, however, four-layer NNs are preferable in general, due to the smaller numbers
of qualifier candidates generated.
Three-layer NNs can still be a choice, if program or invariant synthesis tools (that use our tool as a backend for finding qualifiers)
work well with a very large number of candidate qualifiers; for example, the
ICE-learning-based CHC solver
\hoice{}~\cite{DBLP:journals/jar/ChampionCKS20} seems to
work well with 100 candidate qualifiers (but probably not for 1000 candidates).

As for the 32:32 four-layer NN (which performed best among four-layer NNs),
only 5 correct qualifiers were missed in total and all of them came from Instance \#10 shown on the right-hand side
of Figure~\ref{fig:data2d}. Among the four qualifiers, \(x-y-9>0\) was missed in all the three runs,
and \(x+y>0\) was missed in two of the three runs; this is somewhat expected, as it is quite subtle
to recognize the line \(x-y-9=0\) also for a human being. The NN instead found the qualifiers like \(x>6\) and \(y<-5\)
to separate the positive and negative examples.
\iffull
We were rather surprised that the NN could recognize
the correct qualifiers for other subtle instances like \#2 and \#13 in Appendix~\ref{sec:2d4instances}.
\fi

Figure~\ref{fig:weight-qualifiers} shows the effect of the prioritization of qualifiers, as discussed in
Section~\ref{sec:method}.
We have sorted the hidden nodes in the second layer based on
the weights of the third and fourth layers,
and visualized, in the graph on the left-hand side, the average number of
correct qualifiers (per run)
extracted from top 50\%, 75\%, and 100\% of the hidden nodes.
The graph on the right-hand side shows the average number of candidate qualifiers (per run)
 extracted from top 50\%, 75\%, and 100\% of the hidden nodes.
 As can be seen in the graphs, while the number of candidate qualifiers is almost linear in
 the number of hidden nodes considered, most of the correct qualifiers were found from top 75\% of
 the hidden nodes. This justifies our strategy to prioritize qualifier candidates based on the weights of
 the third and fourth layers.

 Appendix~\ref{sec:act} also reports experimental results to compare
 the sigmoid function with other activation functions.
\begin{figure}
  \begin{center}
    \includegraphics[scale=0.4]{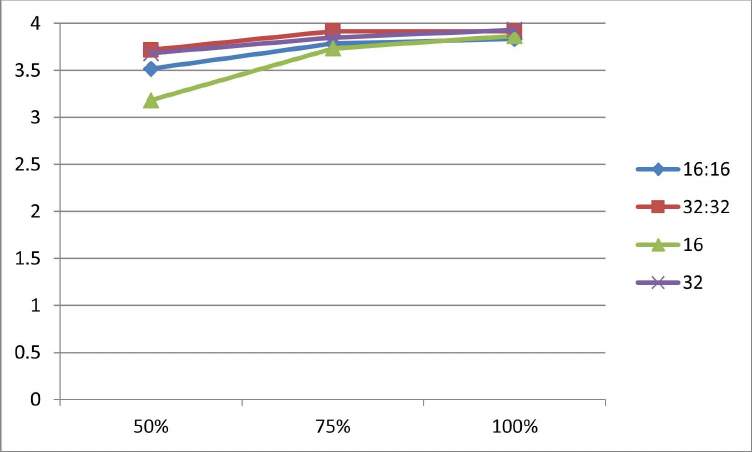}
    \includegraphics[scale=0.4]{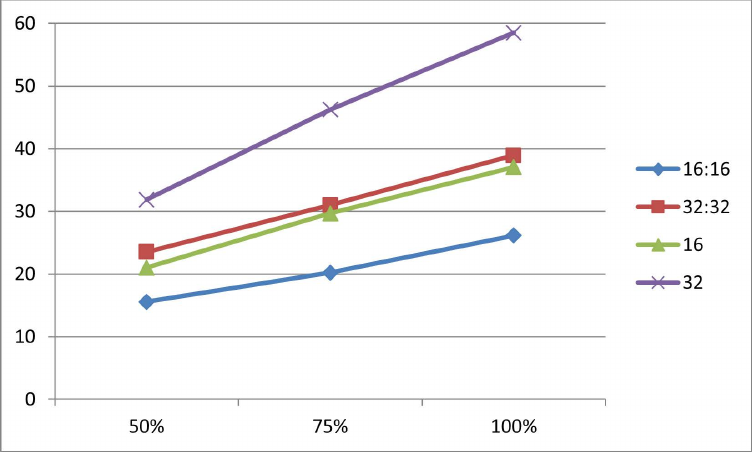}
    \end{center}
  \caption{The number of correct qualifiers discovered (left) and
    the number of qualifier candidates generated (right).
  Most of the correct qualifiers appear in top 50\% of the candidates.}
  \label{fig:weight-qualifiers}
\end{figure}

To check whether our method scales to larger formulas,
we have also tested our tool for formulas of the form
\((A\land B\land C)\lor (D\land E\land F)\lor (G\land H\land I)\).
where \(A,\ldots,I\) are linear inequalities of the form \(a x+ b y + c>0\), and
\(a\) and \(b\) are integers such that
\(-4 \le a, b\le 4\) and \(-30\le c\le 30\) with \(ab\ne 0\).
We set \(P = \set{(x,y)\in \Z^2 \cap ([-30,30]\times[-30,30]) \mid \models \form(x,y)}\)
and \(N = \set{(x,y)\in \Z^2 \cap ([-30,30]\times[-30,30]) \mid \models \neg\form(x,y)}\)
as the sets of positive and negative examples, respectively.
We have collected 20 problem instances, by randomly generating
formulas and then manually filtering out those that
have a simpler representation than
\((A\land B\land C)\lor (D\land E\land F)\lor (G\land H\land I)\).
Table~\ref{tab:learn-2d9} summarizes the experimental results.
We have used the log function as the loss function.
For the qualifier extraction,
we extracted five (as opposed to three, in the experiments above) inequalities of the form
(\(c_0+c_1x_1+\cdots +c_kx_k > e\) for \(e\in\set{-2,-1,0,1,2}\))
from each hidden node. The threshold for the loss was set to \(10^{-5}\),
except that it was set to \(10^{-4}\)
for the three-layer NN with 32 nodes.
\begin{table}[tbp]
\caption{Experimental Results for Learning Formulas of the form \((A\land B\land C)\lor (D\land E\land F)\lor (G\land H\land I)\).}
\label{tab:learn-2d9}
  \begin{center}
\begin{tabular}{|c|c|c|c|c|c|}
\hline   
\#hidden nodes & \#retry & \%success & \%qualifiers & \#candidates
& time (sec.)\\
\hline
\hline
32 &37 & 83.3\% & 86.5\% & 126.6 & 78.7\\
\hline
32:4 & 13 & 78.3\% & 84.6\% & 127.3 & 79.6\\
\hline
32:32 & 0& 70.0\% & 79.6\% & 134.1 & 71.1\\
\hline
64 &1 & 100\% & 90.2\% & 126.6 & 200.1\\
\hline
64:16 & 0& 96.7\% & 86.7\% & 204.9 & 86.1\\
\hline
64:64 & 0& 91.7\% & 84.1\% & 215.3 & 107.6\\
\hline
\end{tabular}
\end{center}
\end{table}

The result indicates that our method works reasonably well even for 
this case, if hyper-parameters (especially, the number of nodes) are chosen
appropriately; how to adjust the hyper-parameters is left for future work.
Interestingly, the result tends to be better for NNs with
a smaller number of nodes in the third layer.
Our rationale for this is that 
a smaller number of nodes in the third layer forces the nodes
in the second layer more strongly to recognize appropriate features.

\subsection{Extensions for Non-Linear Constraints}
\label{sec:nonlinear}
We can easily extend our approach to synthesize formulas
consisting of non-linear constraints such as polynomial constraints of a bounded
degree
and modulo operations. For that purpose, we just need to add auxiliary inputs
like \(x^2, x\,\mod\,2\) to a NN.
We have tested our tool for quadratic inequalities of the form
\(ax^2+bxy+cy^2+dx+ey+f>0\) (where \(-4\le a,b,c,d,e\le 4\)
and \(-9\le f\le 9\); for ovals, we allowed \(f\) to range over \([-200,199]\)
because there are only few positive examples for small values of \(f\)),
and their disjunctions.
For a quadratic formula \(\form(x,y)\), we set
\(P=\set{(x,y,x^2,xy,y^2)\mid (x,y)\in \Z^2 \cap ([-20,20]\times[-20,20]), \models \form(x,y)}\)
and 
\(N=\set{(x,y,x^2,xy,y^2)\mid (x,y)\in \Z^2 \cap ([-20,20]\times[-20,20]), \models \neg\form(x,y)}\) as the sets of positive and negative examples, respectively.

The table below shows the result for a single quadratic inequality.
\iffull
We have prepared four instances for each of ovals, parabolas, and hyperbolas;
the instances used in the experiment
are visualized in Appendix~\ref{sec:deg2instances} (Instances \#1--\#12).
\fi
As can be seen in the table, the tool worked very well.
\begin{center}
\begin{tabular}{|c|c|c|c|c|c|c|}
\hline   
\#hidden nodes & loss func. & \#retry & \%success & \%qualifiers & \#candidates & time (sec.)\\
\hline
4 & log & 0 & 100\% & 100\% & 7.4 & 22.2\\
\hline
8 & log & 0 & 100\% & 100\% & 10.7 & 18.3\\
\hline
\end{tabular}
\end{center}

\iffull
We have also prepared six instances of formulas of the form \(A\lor B\),
where \(A\) and \(B\) are quadratic inequalities; they are visualized as
Instances \#13--\#18 in Appendix~\ref{sec:deg2instances}.
\fi
The result is shown in the table below.
The NN with 32 nodes performed reasonably well, considering the difficulty
of the task. 
All the failures actually came from Instances \#16 and \#18, and the tool succeeded
for the other instances.
For \#16, the tool struggled to correctly recognize the oval
\(-2x^2-2xy-2y^2-3x-4y+199>0\) (\(-2x^2-2xy-2y^2-3x-4y+197>0\) was instead
generated as a candidate; they differ at only two points \((0,9)\) and \((0,-11)\)),
and for \#18, it failed to recognize the hyperbola \(-4x^2-4xy+y^2+2x-2y-2>0\).
\begin{center}
\begin{tabular}{|c|c|c|c|c|c|c|}
\hline   
\#hidden nodes & loss func. & \#retry & \%success & \%qualifiers & \#candidates & time (sec.)\\
\hline
8 & log & 5 & 50\% & 52.8\% & 22.1 & 43.9\\
\hline
16 & log & 0 & 44.4\% & 55.6\% & 38.6 & 26.1\\
\hline
32 & log & 0 & 66.7\% & 80.6\% & 72.1 & 25.4\\
\hline
\end{tabular}
\end{center}

\section{Predicate Synthesis from Implication Constraints and its Application to CHC Solving}
\label{sec:ICE}

We have so far considered the synthesis of logical formulas
in the classical setting of supervised learning of classification,
where positive and negative examples are given. 
In the context of program verification, we are also given
so called \emph{implication constraints}~\cite{garg_2014,DBLP:journals/jar/ChampionCKS20},
like \(p(1) \imp p(2)\), which means ``if $p(1)$ is a positive example, so is $p(2)$''
(but we do not know whether $p(1)$ is a positive or negative example).
As advocated by Garg et al.~\cite{garg_2014},
implication constraints play an important role in the discovery of
an \emph{inductive} invariant
(i.e. an invariant \(\mathit{Inv}\) that satisfies a certain condition of
the form \(\forall x,y.(\psi(\Inv,x,y) \imp \Inv(x))\);
for a state transition system, \(\psi(\Inv,x,y)\) is of the form
\(\Inv(y)\land R(y,x)\) where \(R\) is the transition relation).
As discussed below, our framework of NN-guided synthesis can easily be adapted to
deal with implication constraints.
Implication constraints are also called \emph{implication examples} below.

\subsection{The Extended Synthesis Problem and Our Method}
We first define the extended synthesis problem.
A \emph{predicate signature} is
 a map from a finite set
 \(\PV=\set{p_1,\ldots,p_m}\) of predicate variables to the set \(\Nat\) of natural numbers.
 For a predicate signature \(\sig\),
 \(\sig(p_i)\) denotes the arity of the predicate
\(p_i\). We call a tuple of the form \((p_i, n_1,\ldots,n_{\sig(p_i)})\) with
\(n_j\in \Z\) an \emph{atom}, and often write \(p_i(n_1,\ldots,n_{\sig(p_i)})\) for it;
we also write \(\seq{n}\)
for a sequence \(n_1,\ldots,n_{\sig(p_i)}\) and write \(p_i(\seq{n})\) for
the atom \(p_i(n_1,\ldots,n_{\sig(p_i)})\). We write \(\Atoms{\sig}\) for the
set of atoms consisting of the predicates given by the signature \(\sig\).
\begin{definition}[Predicate synthesis problem with Implication Examples]
  The goal of the
  predicate synthesis problem with positive/negative/implication examples
  (the PNI synthesis problem, for short) is, given:
  \begin{enumerate}
\item a signature \(\sig \in \PV\to \Nat\); and 
  \item a set \(I\) of \emph{implication examples} of the form
  \(a_1\land \cdots \land a_k\imp b_1\lor \cdots \lor b_\ell\) where
  \(a_1,\ldots,a_k,b_1,\ldots,b_\ell\in \Atoms{\sig}\), and \(k+\ell>0\)
  (but \(k\) or \(\ell\) may be \(0\)) 
  \end{enumerate}
as an input, 
to find a map \(\theta\) that assigns, to each \(p_i\in \PV\),
a logical formula \(\form_i(x_1,\ldots,x_{\sig(p_i)})\) such that
\(\models \theta a_1\land \cdots \land \theta a_k\imp
  \theta b_1\lor \cdots \lor \theta b_\ell\)
  holds for each implication example
  \(a_1\land \cdots \land a_k\imp b_1\lor \cdots \lor b_\ell\in I\).
Here, for an atom \(a=p(n_1,\ldots,n_j)\), \(\theta a\) is defined as
\((\theta p)[n_1/x_1,\ldots,n_j/x_j]\).
We call an implication example of the form \(\imp b\)
(\(a\imp\), resp.)
as a positive (negative, resp.) example,
and write \(P\subseteq I\) and \(N\subseteq I\) for the sets of positive and
negative examples
respectively.
\end{definition}

\begin{example}
  Let \(\sig = \set{p\mapsto 1, q\mapsto 1}\)
  and \(I\) be \(P\cup N\cup I'\) where:
  \[
  \begin{array}{l}
    P = \set{\imp p(0), \imp q(1)}\qquad
    N = \set{p(1)\imp, q(0)\imp}\\
    I' = \{p(2)\imp q(3), q(3)\imp p(4),
      p(2)\land q(2)\imp, p(3)\land q(3)\imp,\\\qquad
      p(4)\land q(4)\imp,
      \imp p(2)\lor q(2),       \imp p(3)\lor q(3),   \imp p(4)\lor q(4)\}
  \end{array}
  \]
  Then \(\theta = \set{p\mapsto x_1\mod 2=0, q\mapsto x_1\mod 2=1}\)
  is a valid solution for the synthesis problem \((\sig, I)\). \qed
\end{example}


We generalize the method described in Section~\ref{sec:method} as follows.

\begin{asparaenum}
\item Prepare a NN \(\N_i\) for \emph{each} predicate variable \(p_i\).
  \(\N_i\) has \(\sig(p_i)\) input nodes (plus additional inputs, if
  we aim to generate non-linear formulas, as discussed in Section~\ref{sec:nonlinear}).
  For an atom \(a\equiv p_i(\seq{n})\), we write \(o_a\) for the output of \(\N_i\) for
  \(\seq{n}\) below (note that
  the value of \(o_a\) changes during the course of training).
\item Train all the NNs \(\N_1,\ldots,\N_m\) together.
  For each atom \(p_i(\seq{n})\) occurring in implication examples,
  \(\seq{n}\) is used as a training datum for \(\N_i\).
  For each implication example
  \(e \equiv a_1\land \cdots \land a_k\imp b_1\lor \cdots \lor b_\ell\in I\),
  we define the loss \(\loss{e}\) for the example by:
  \[
  \loss{e} =
  -\log (1-\prod_{i=1}^k o_{a_i}\cdot \prod_{j=1}^\ell (1-o_{b_j})).\footnote{
    In the implementation,
    we approximated
    \(\loss{e}\) by
    \(-\log \max(\epsilon,1-\prod_{i=1}^k o_{a_i}\cdot \prod_{j=1}^\ell (1-o_{b_j}))\)
    for a small positive number \(\epsilon\)
     in order to avoid an overflow of the floating point arithmetic.}
  \]
  The idea is to ensure that the value of \(\loss{e}\)
  is \(0\) if one of \(o_{a_i}\)'s is \(0\) or
  one of \(o_{b_j}\)'s is \(1\), and that the value of \(\loss{e}\) is
  positive otherwise.
  This reflects the fact that
  \(a_1\land \cdots \land a_k\imp b_1\lor \cdots \lor b_\ell\)
  holds just if one of \(a_i\)'s is false or one of \(b_j\)'s is true.
  Note that the case where \(k=0\) and \(\ell=1\) (\(k=1\) and \(\ell=0\), resp.)
  coincides with the logarithmic loss function for positive (negative, resp.) examples
  in Section~\ref{sec:method}.
  Set the overall loss of the current NNs as
  the average of \(\loss{e}\) among all the implication constraints,
  and use the gradient descent to update the weights and biases of NNs.
  Repeat the training until all the implication constraints
  are satisfied (by considering values greater than \(0.5\) as true,
  and those less than \(0.5\) as false).
\item Extract a set \(Q_i\) of qualifiers from each trained \(\N_i\),
  as in Section~\ref{sec:method}.
\item Synthesize a formula for the predicate \(p_i\) as a Boolean combination of \(Q_i\).
  This phase is also the same as Section~\ref{sec:method}, except that,
  as the label \(\ell_a\) for each atom \(a\), we use the prediction of
  the trained NNs. Note that unlike in the setting of the previous section
  where we had only positive and negative examples, 
  we may not know the correct label of some atom due to the existence of
  implication examples. For example, given \(p(0)\lor p(1)\imp\) and \(\imp
  p(0)\lor p(1)\), we do not know which of \(p(0)\) and \(p(1)\) should hold.
  We trust the output of the trained NN in such a case. Since we have
  trained the NNs until all the implication constraints are satisfied,
  it is guaranteed that, for positive and negative examples,
  the outputs of NNs match the correct labels.
  Thus, the overall method strictly subsumes the one described in Section~\ref{sec:method}.
\end{asparaenum}

\subsection{Preliminary Experiments}
\label{sec:iceex}
We have extended the tool based on the method described above, and tested it
for several examples. We report some of them below.

As a stress test for implication constraints, 
we have prepared the following input, with no positive/negative examples
(where \(\sig = \set{p \mapsto 1}\)):
\[
\begin{array}{l}
I = \set{p(2n)\land p(2n+1)\imp \;\mid n\in [-10,10]} \\\quad
\cup \set{\imp p(2n)\lor p(2n+1) \;\mid n\in [-10,10]}
\end{array}
\]
The implication examples mean that, for each integer \(n\in [-10,10]\), exactly one of
\(p(2n)\) and \(p(2n+1)\) is true.
We ran our tool with an option to enable the ``mod 2'' operation,
and obtained \(\set{p\mapsto x_1\mod 2<1}\) as a solution (which is correct;
another solution is \(x_1\mod 2>0\)).

\begin{figure}[tbp]
  \begin{center}
    \includegraphics[scale=0.5]{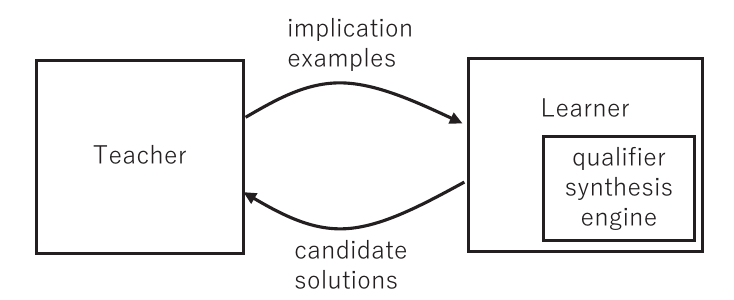}
    \vspace*{-.4cm}
  \end{center}
  \caption{Inside \hoice{}}
  \label{fig:hoice}
\end{figure}

We have also tested our tool for several instances of the 
CHC (Constrained Horn Clauses) solving problem~\cite{Bjorner15}.
A constrained Horn clause is a formula of the form
\(A_1\land \cdots \land A_k\imp A_0\), where
each \(A_i\) is a constraint (of linear arithmetic in this paper)
or an atomic formula of the form \(p(e_1,\ldots,e_k)\) where \(p\) is a predicate
variable. The goal of CHC solving is, given a set of constrained Horn clauses,
to check whether there is a substitution for predicate variables that makes
all the clauses valid (and if so, output a substitution).
Various program verification
problems~\cite{Bjorner15}, as well as the problem of
finding inductive invariants, can be reduced to the problem of CHC solving.
Various CHC solvers~\cite{DBLP:journals/fmsd/KomuravelliGC16,Eldarica,DBLP:conf/aaai/Satake0Y20,DBLP:journals/jar/ChampionCKS20} have been implemented so far,
and \hoice{}~\cite{DBLP:journals/jar/ChampionCKS20} is
one of the state-of-the-art solvers, which is based on the ICE-learning
framework.
As illustrated in Figure~\ref{fig:hoice}, \hoice{} consists of two main components:
a teacher and a learner.
The learner generates a candidate solution (a map from predicate variables
to formulas) from implication examples, and the teacher checks whether the candidate solution is valid
(i.e. satisfies all the clauses) by calling SMT solvers, and if not, generates a new implication example.
The learner has a qualifier synthesis engine and combines it with a method for Boolean decision tree
construction~\cite{DBLP:conf/popl/0001NMR16}. 
The main bottleneck of \hoice{} has been the qualifier synthesis engine,
and the aim of our preliminary experiments reported below
is thus to check whether our NN-guided synthesis method can be used to reduce the bottleneck.

To evaluate the usefulness of our NN-guided synthesis, we have implemented the following
procedure, using \hoice{} and our NN-guided synthesis tool (called \NNG{}) as backends.
\begin{itemize}
\item[Step 1:] Run \hoice{} for 10 seconds to collect implication examples.
\item[Step 2:] Run \NNG{} to learn qualifiers. 
\item[Step 3:] Re-run \hoice{} with the qualifiers as hints (where the time
  limit is initially set to 10 seconds).
\item[Step 4:] If Step~3 fails, collect implication examples from the execution of
  Step~3,  and go back to Step 2, with the time limit for \hoice{} increased by 5 seconds.
\end{itemize}
In Step~2, we used a four-layer NN, and set the numbers of hidden nodes in
the second and third layers to 4. When \NNG{} returns a formula,
the inequality constraints that constitute the formula as qualifiers are passed to
Step~3; otherwise, all the qualifier candidates extracted from the trained NNs
are passed to Step~3.

We have collected the following CHC problems that plain \hoice{} cannot solve. 
\begin{asparaitem}
\item Two problems that arose during our prior analysis of the bottleneck \hoice{} (\texttt{plus}, \texttt{plusminus}).
The problem \texttt{plusminus} consists of the following CHCs.
\newcommand\plus{\mathit{plus}}
\newcommand\minus{\mathit{minus}}
\[
\begin{array}{rclcrcl}
  \plus(m,n,r) \land \minus(r,n,s)&\imp& m=s &\qquad&
  \true &\imp& \plus(m,0,m)\\
  n>0\land \plus(m, n-1, r) &\imp& \plus(m,n,r+2)&&
  \true &\imp& \minus(m,0,m)\\
  n>0\land \minus(m, n-1, r) &\imp& \minus(m,n,r-2)\\
\end{array}
\]
The above CHC is satisfied for 
\(\plus(m,n,r)\equiv r=m+2n,  \minus(m,n,r)\equiv r=m-2n\).
The ICE-based CHC solver \hoice{}~\cite{DBLP:journals/jar/ChampionCKS20},
however, fails to solve the CHC (and runs forever), as
\hoice{} is not very good at finding qualifiers involving more than
two variables.
\item The five problems from
  the Inv Track of SyGus 2018 Competition (\url{https://github.com/SyGuS-Org/benchmarks/tree/master/comp/2018/Inv_Track},\\
 \texttt{cggmp2005\_variant\_true-unreach-call\_true-termination}, \texttt{jmbl\_cggmp-new}, 
  \texttt{fib\_17n}, \texttt{fib\_32}, and \texttt{jmbl\_hola.07}, which are named
  \texttt{cggmp}, \texttt{cggmp-new}, \texttt{fib17}, \texttt{fib32}, and \texttt{hola.07}
  respectively in Table~\ref{tab:expCHC});
  the other problems can be solved by plain \hoice{}, hence excluded out).
\item The problem \#93 from the code2inv benchmark set used in \cite{DBLP:conf/iclr/RyanWYGJ20}
  (93; the other 123 problems  can be solved by plain \hoice{} within 60 seconds).
\item Two problems (\texttt{pldi082\_unbounded1.ml} and \texttt{xyz.ml})
  from the benchmark set (\url{https://github.com/nyu-acsys/drift}) of Drift~\cite{DBLP:journals/pacmpl/PavlinovicSW21},%
  \footnote{The tool \texttt{r\_type} was used to extract CHCs. The source programs have been slightly modified
    to remove Boolean arguments from predicates.} with a variant \texttt{xyz\_v.ml} of
  \texttt{xyz.ml}, obtained by generalizing the initial values of some variables.
\end{asparaitem}
The implementation and the benchmark set described above are
available at \url{https://github.com/naokikob/neugus}.

Table~\ref{tab:expCHC} summarizes the experimental results.
The experiments were conducted on the same machine as those of Section~\ref{sec:method}.
We used \hoice{} 1.8.3 as a backend.
We used 4-layer NNs with 32 and 8 hidden nodes in the second and third layers respectively.
The column '\#pred' shows the number of predicates in the CHC problem,
and columns '\#P', '\#N', and '\#I' respectively show the numbers of
positive, negative, and implication examples  (the maximum numbers in the three runs).
The column 'Cycle' shows the minimum and maximum numbers of Step~2 in the three runs.
The column 'Time' shows the
execution time in seconds (which is dominated by \NNG{}), where
the time is the average for three runs, and '-' indicates a time-out
(with the time limit of 600 seconds). 
The next column shows key qualifiers
found by \NNG{}. For comparison with our combination of \hoice{} and \NNG{},
the last column shows the times (in seconds) spent by 
Z3 (as a CHC solver)
for solving the problems.\footnote{Recall that our benchmark set collects only the problems that plain \hoice{} cannot solve: although many of those problems
can be solved by Z3~\cite{DBLP:journals/fmsd/KomuravelliGC16} much more quickly,
there are also problems that Z3 cannot solve but (plain) \hoice{} can.}

\begin{table}[tbp]
  \label{tab:expCHC}
  \caption{Experimental results on CHC solving}
  \begin{center}
  \begin{tabular}{|l|r|r|r|r|c|c|c|c|}
    \hline
    Problem & \#pred & \#P & \#N & \#I & Cycle & Time & Key qualifiers found 
    & Z3\\
    \hline
    \texttt{plus}&1 & 29  & 48 & 35 & 1  & 23.8 & \(x_0+2x_1+x_2=0\)  & -\\
    \hline
    \texttt{plusminus}&2 & 90 & 137 & 42 & 2 & 63.2 & \(x_0+2x_1-x_1=0\)
    & -\\
    \hline
    \texttt{cggmp}&1 & 25 & 95 & 11 & 1 & 20.0 & \(x_0+2x_1-x_2=0\)  & 0.05\\
    \hline
    \texttt{cggmp-new}&1 & 8 & 97 & 15 & 1 & 24.1 & \(x_0+2x_1\ge 40, x_0+2x_1\le 42\)  & 0.12\\
    \hline
    \texttt{fib17n}&1 & 44 & 139 & 35 & 7--10 & 349.7 & \(-x_0+x_2-x_3\ge 0\)  & 0.14\\
    \hline
    \texttt{fib32}&1 & 42 & 222 & 17  & -& - & -   & -\\
    \hline
    \texttt{fib32 -mod2}&1 & 42 & 222 & 17  & 1--3 & 53.7 & \(x_3\mod 2=0\)   & -\\
    \hline
    \texttt{hola.07}&1 & 78& 314 & 12& 2 -- 3  & 72.5 & \(x_0+x_1=3x_2\) & 0.05 \\
    \hline
    \texttt{codeinv93}&1 & 32  & 260 & 23 & 2 -- 3 & 73.2 &  \(-3x_0+x_2+x_3=0\)  & 0.05\\
    \hline
    \texttt{pldi082}&2 &  15 & 42 & 34 & 1 & 24.6 & \(2x_0-x_1-x_2\ge -2\)   & -\\
    \hline
    \texttt{xyz}&2 & 56 & 6 & 142  & -&- &-   & 0.33\\
    \hline
    \texttt{xyz -gen}&2 & 251 & 139 & 0  & 3 -- 5& 185.9 &\(x_0 + 2x_1  \ge 0\)   & 0.33\\
    \hline
    \texttt{xyz\_v}&2 & 160 & 303 & 43 & 2 -- 8& 217.3 &
    \(x_0-x_1-2x_2+2x_3\ge 0\)   & -\\
    \hline
  \end{tabular}
  \end{center}
\end{table}

The results reported above indicate that our tool can indeed be used to improve the
qualifier discovery engine of ICE-learning-based CHC solver \hoice{}, which has been the 
main bottleneck of \hoice{}.
With the default option, \NNG{} timed out for \texttt{fib32}
and \texttt{xyz}.
For the problem \texttt{fib32}, the ``mod 2'' constraint is required.
The row ``fib32 -mod2'' shows the result obtained by running \NNG{} with
the ``mod 2'' constraint enabled.
For the problem \texttt{xyz}, we discovered that the main obstacle was on the side of \hoice{}:
 \hoice{} (in the first step)
 finds no negative constraint for one of the two predicates,
 and no positive constraint for the other predicate;
 thus, \NNG{} returns ``true'' for the former predicate and ``false'' for the latter predicate.
 We have therefore prepared an option to eagerly collect learning data by applying
 unit-propagation to CHCs; the row ``\texttt{xyz -gen}'' shows the result for this
 option; see Appendix~\ref{app:sec3} for more
detailed analysis of the problem \texttt{xyz}.
Larger experiments on the application to CHC solvers are left for future work.

\iffull
\section{Another Application: Programming with Oracles}
\label{sec:oracle}
As another potential application of our framework of NN-guided synthesis of
logical formulas, we propose a framework of \emph{programming with oracles}.
The overall flow of the framework is shown in
Figure~\ref{fig:oracle}.
\begin{figure}[tbp]
\includegraphics[scale=0.5]{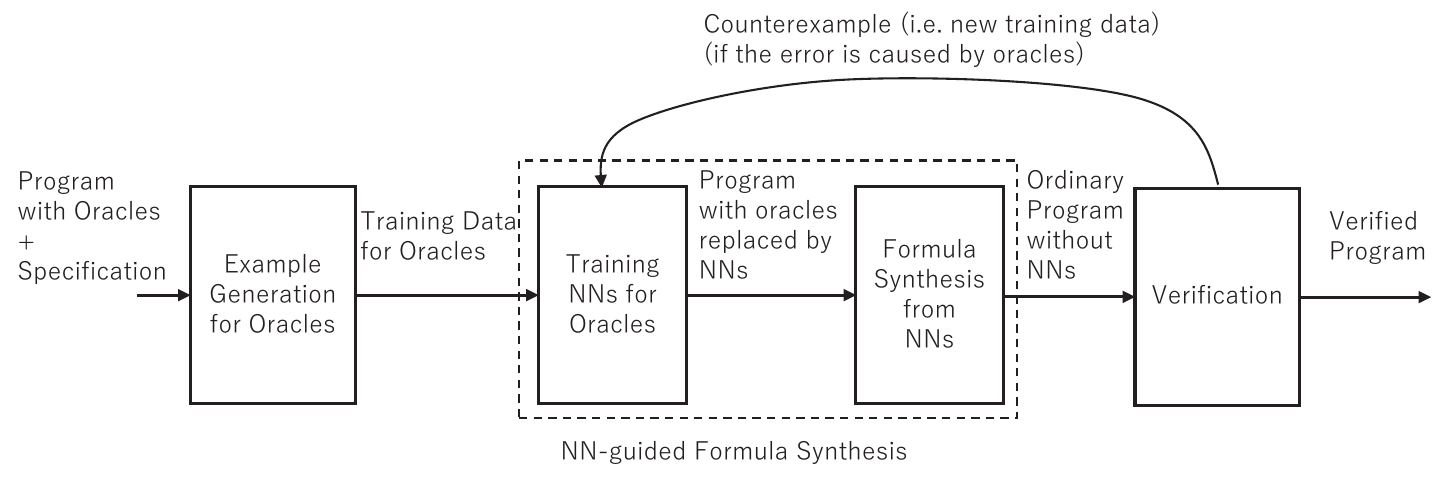}  
\caption{Programming with Oracles}
\label{fig:oracle}
\end{figure}

A programmer provides a program containing ``oracles'', along with
the specification of the entire program, and the goal of the overall
synthesis is to replace the oracles with appropriate code.
Here is an example of a program with oracles and a specification:
\begin{verbatim}
let abs n = if oracle(n) then n else -n
let main n = assert(abs(n)>=0) (* specification of abs *)
\end{verbatim}
Here, the code for function \texttt{abs} contains an oracle,
and the \texttt{main} function gives a specification of \texttt{abs}
by using assertions (which says that for any integer \(n\), \(\texttt{abs}(n)\ge 0\)
should hold). The specification could alternatively be
given in terms of example input/output pairs or pre/post conditions.
(The specification could also be a soft one, like ``minimize the running time''.)
The goal here is to replace \texttt{oracle(n)} with
appropriate code, like \texttt{n>0}.
Thus, the framework can be viewed as a variation of
 the program synthesis from sketch~\cite{solar2008program}, and
the main new ingredient is the use of NNs to guide the synthesis.
In the discussion below,
we assume that oracles are restricted to those that return Boolean values.
(To lift the restriction, we need to extended our formula synthesis method in
Section~\ref{sec:PN} to synthesize arbitrary expressions.)

Each step in Figure~\ref{fig:oracle} is described as follows.
\begin{asparaenum}
\item Example generation: For each oracle in the program,
  we generate positive/negative/implication examples on oracles.
  This can be achieved by first replacing oracles with non-deterministic
  choice, and using random testing or model checking to obtain error traces
  (sequences of oracle calls) that violate the specification.
  For the \texttt{abs} program above,
  we can easily find that any of the following oracle calls leads to
  an assertion failure:
  \[
  \begin{array}{l}
  \oracle(-1)=\true,\ \oracle(-2)=\true,\ \ldots\\
  \oracle(1)=\false,\ \oracle(2)=\false,\ \ldots
  \end{array}
  \]
  Thus, we can set \(P=\set{\imp \oracle(1),\ \imp\oracle(2),\ \ldots}\)
  and \\\(N=\set{\oracle(-1){\imp,}\ \oracle(-2)\imp,\ldots}\) as positive
  and negative examples respectively.

  In general, a single error trace may contain multiple calls of
  multiple oracles,
  like
  \[
  \oracle_1(d_1)=\true; \oracle_2(d_2)=\false; \oracle_1(d_3)=\true,
  \]
  when a program contains multiple kinds of oracles and calls of them occur inside
  a recursion or loop. For the sequence of oracle calls above, we generate
  the following implication example:
  \[
   \neg \oracle_1(d_1) \lor \oracle_2(d_2) \lor \neg \oracle_1(d_3)
   \]
   (or, \(\oracle_1(d_1)\land \oracle_1(d_3)\imp \oracle_2(d_2)\)
   in the syntax of implication examples in the previous section),
   which expresses that at least one of the return values of
   oracle calls is wrong.

   One may also extract examples from successful traces.
   For example, given a 
   successful trace
   \[\oracle_1(d_4)=\true; \oracle_2(d_5)=\true,\]
   we can add \(\oracle_2(d_5)\imp \oracle_1(d_4)\)
   and 
   \(\oracle_1(d_4)\imp \oracle_2(d_5)\)  as
   implication examples.
   The reason why we have the conditions
   \(\oracle_2(d_5)\imp\cdots \) and \(\oracle_1(d_4)\imp \cdots\) is
   that it may be the case that both \(\oracle_1(d_3)=\true\) and
   \(\oracle_2(d_5)=\true\) were wrong but that the whole program execution
   succeeded just by coincidence.
   In general, from a successful trace:
 \[ \oracle_1(d_1)=b_1; \oracle_2(d_2)=b_2; \cdots; \oracle_k(d_k)=b_k,\]
 we add 
 \[ \bigwedge_{j\ne i} (\oracle_j(d_j)=b_j) \imp \oracle_i(d_i)=b_i\]
 for each \(i\in\set{1,\ldots,k}\) as implication constraints.
   In general, however, we need to treat the examples generated from successful traces
   as \emph{soft} constraints, to avoid a contradiction.
   For example, consider the following program:
\begin{verbatim}
let f x = if oracle(x) then 0 else 0
let main n = assert(f n=0)
\end{verbatim}
In this case, the return value of \verb|oracle| does not matter.
Since both \(\oracle(0)=\true\) and \(\oracle(0)=\false\)
are successful traces, we get \(\imp \oracle(0)\) and \(\oracle(0)\imp\),
which cannot be simultaneously satisfied; hence, they should
be treated as soft constraints.
   The distinction between hard and
   soft constraints can be reflected in the NN training phase
   by adjusting the loss function (e.g. by giving small weights to the loss
   caused by soft constraints).
 \item NN training: just as described in Section~\ref{sec:ICE}.
   Alternatively, one could directly verify NNs~\cite{DBLP:conf/sp/GehrMDTCV18,zhao2020learning,DBLP:conf/nips/AndersonVDC20,DBLP:conf/cav/PulinaT10,DBLP:conf/aaai/NarodytskaKRSW18} and use them as oracles,
   or ``shield'' NNs by ordinary code~\cite{DBLP:conf/pldi/ZhuXMJ19},
 without proceeding to the next
   phase for synthesis;
   these are also interesting directions, but outside the scope of the present paper.
 \item Formula synthesis: this phase is also as described in Section~\ref{sec:ICE}.
 \item Verification:
   Oracles are replaced with the formulas synthesized in the previous phase,
   and the resulting programs are verified and/or tested.
   If an error trace is found, generate implication constraints from it
   and go back to the training phase. If there is no oracle call sequence that
   avoids the error trace, then a bug is on the side of the program with oracles,
   hence report it to the programmer.
\end{asparaenum}

Our method for the NN-guided synthesis would serve as a key ingredient in the framework
above. We have not yet implemented the framework, which is left for future work.
To test the feasibility of the proposed framework, we have
tested the procedure above for some benchmark programs of higher-order
program verification\footnote{https://github.com/hopv/benchmarks.},
by manually emulating the steps outside our NN-guided synthesis tool.

\iffull
Among the most non-trivial examples was the synthesis of the condition of McCarthy's 91 function:
\begin{verbatim}
let rec mc91 x = if oracle(x) then x - 10 else mc91 (mc91 (x + 11))
let main n = if n <= 101 then assert (mc91 n = 91)
\end{verbatim}
We could successfully synthesize the formula \(x>100\) for \verb|oracle(x)| above.
More details are reported in Appendix~\ref{sec:oracle-exp}.
\fi

\fi
\section{Related Work}
\label{sec:rel}

There have recently been a number of studies on verification of
neural networks~\cite{DBLP:conf/sp/GehrMDTCV18,zhao2020learning,DBLP:conf/nips/AndersonVDC20,DBLP:conf/cav/PulinaT10,DBLP:conf/aaai/NarodytskaKRSW18}:
see \cite{DBLP:conf/cav/HuangKWW17} for an extensive survey.
In contrast, the end goal of the present paper is
to apply neural networks to verification and synthesis of \emph{classical}
programs.
Closest to our use of NNs is the work of Ryan et al.~\cite{DBLP:conf/iclr/RyanWYGJ20,DBLP:conf/pldi/YaoRWJG20}
on Continuous Logic Network (CLN). CLN is a special neural network that imitates a logical formula
(analogously like symbolic regressions discussed below),
and it was applied to learn loop invariants. The main differences are:
(i) The shape of a formula must be fixed in their original approach~\cite{DBLP:conf/iclr/RyanWYGJ20}, while
it need not in our method, thanks to our hybrid approach of extracting just qualifiers and using a classical
method for constructing its Boolean combinations. Although their later approach~\cite{DBLP:conf/pldi/YaoRWJG20}
relaxes the shape restriction, it still seems less flexible than ours (in fact, the shape of invariants found by their tools seem limited,
according to the code available at
\url{https://github.com/gryan11/cln2inv} and
\url{https://github.com/jyao15/G-CLN}).
(ii) We consider a more general learning problem, using not only positive examples, but also negative and
implication examples. This is important for applications to CHC solving discussed in Section~\ref{sec:ICE}
and oracle-based program synthesis
\iffull in Appendix~\ref{sec:oracle}.
\fi



Finding inductive invariants has been the main bottleneck of
program verification, both for automated tools (where tools have
to automatically find invariants) and semi-automated tools (where users
have to provide invariants as hints for verification).
In particular, finding appropriate qualifiers (sometimes
 called \emph{features}~\cite{DBLP:conf/pldi/PadhiSM16},
 and also \emph{predicates} in verification methods
 based on predicate abstraction), which are atomic
formulas that constitute inductive invariants, has been a challenge.
Various machine learning techniques have recently been applied to
the discovery of invariants and/or qualifiers. As already mentioned,
Garg et al.~\cite{garg_2014,DBLP:conf/popl/0001NMR16}
proposed a framework of semi-supervised learning called
ICE learning, where implication constraints are provided to a learner in addition
to positive and negative examples. The framework has later been generalized for
CHC solving~\cite{DBLP:journals/jar/ChampionCKS20,DBLP:journals/pacmpl/EzudheenND0M18},
but the discovery of qualifiers remained as a main bottleneck.

\begin{figure}[tbp]
  \begin{center}
    \includegraphics[scale=0.4]{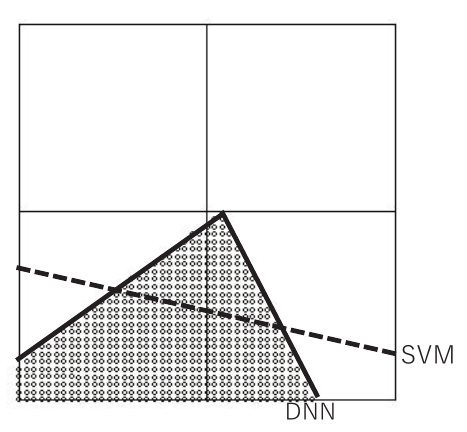}
    \vspace*{-.3cm}
    \end{center}
  \caption{SVM vs NN: Small circles denote positive examples,
    and the white space is filled with negative examples.
    The dashed line shows a linear classifier typically
    output by a SVM, while the thick line shows the (complete) classifier output by
    a NN.}
    \label{fig:svm}
\end{figure}

To address the issue of qualifier discovery, Zhu et al.~\cite{DBLP:conf/pldi/ZhuMJ18}
proposed a use of SVMs (support vector machines). Whilst
SVMs are typically much faster than NNs, there are significant
shortcomings: (i) SVMs are not good at finding a Boolean combination of linear inequalities
(like \(A\land B\)) as a classifier.
To address the issue, they combined SVMs (to find each qualifier \(A,B\))
with the Boolean decision tree construction~\cite{DBLP:conf/popl/0001NMR16},
but it is in general unlikely that SVMs generate \(A\) and/or \(B\) as classifiers
when \(A\land B\) is a complete classifier (see Figure~\ref{fig:svm}).
(ii) SVMs do not properly take implication constraints into account.
Zhu et al.~\cite{DBLP:conf/pldi/ZhuMJ18}
label the data occurring only in implication constraints
as positive or negative examples in an ad hoc manner,
and pass the labeled data to SVMs. The problem with that approach is
that the labeling of data is performed
without considering the overall classification problem.
Recall the example problem in Section~\ref{sec:iceex} consisting of
implication constraints of the form
\(p(2n)\land p(2n+1)\imp\) and \(\imp p(2n)\lor p(2n+1)\) for \(n\in[-10,10]\).
In this case, there are \(2^{21}\) possible ways to classify data,
of which only two classifications (\(p(2n)=\true\) and \(p(2n+1)=\false\) for all \(n\),
or \(p(2n+1)=\true\) and \(p(2n)=\false\) for all \(n\)) lead to
a concise classification of \(n\mod 2=0\) or \(n\mod 2=1\).
Sharma et al.~\cite{DBLP:conf/cav/SharmaNA12} also applied SVMs to find interpolants.
It would be interesting to investigate whether our approach is also useful for
finding interpolants.

Padhi et al.~\cite{DBLP:conf/pldi/PadhiSM16,DBLP:conf/cav/PadhiMN019}
proposed a method for finding qualifiers (features, in their terminology)
based on a technique of syntax-guided program synthesis
and combined it with Boolean function learning.
Since they enumerate possible features in the increasing order of
the size of feature expressions (\cite{DBLP:conf/pldi/PadhiSM16}, Figure~5),
we think they are not good at finding a feature of large size (like
quadratic constraints used in our experiment in Section~\ref{sec:nonlinear}).
Anyway, since our technique has its own defects (in particular, for finding simple
invariants, it is too much slower than the techniques above), our technique
should be used as complementary to existing techniques.
Si et al.~\cite{DBLP:conf/nips/SiDRNS18} used neural reinforcement learning
for learning loop invariants, but in a way quite different from ours.
Rather than finding invariants in a data-driven manner, they
let NNs to learn invariants from a graph representation of programs.

Although the goal is different, our technique
is also related with the NN-based symbolic regression or
extrapolation~\cite{DBLP:conf/iclr/MartiusL17,Petersen21ICLR,pmlr-v80-sahoo18a}, where the goal is to learn a simple 
mathematical function expression (like \(f(x)=\sin(x+x^2)\))
from sample pairs of inputs and outputs.
To this end, Martius and Lambert's method~\cite{DBLP:conf/iclr/MartiusL17}
prepares a NN whose activation functions are basic mathematical functions (like
\(\sin\)), trains it, and extracts the function from the trained NN.
The main difference of our approach from theirs is to use NNs as a gray (rather than white)
box, only for extracting qualifiers, rather than extracting the whole function computed
by NN.
Nevertheless, we expect that
some of their techniques would be useful also in our context,
especially for learning non-linear qualifiers.

\section{Conclusion}
\label{sec:conc}
We have proposed a novel method for synthesizing logical formulas
using neural networks as \emph{gray} boxes.
\iffull
Our key insight was that we should be able to extract useful information
from trained NNs, if the NNs are suitably designed with the extraction in mind.
\fi
The results of our preliminary experiments are quite promising.
We have also discussed an application of our NN-guided synthesis to
program verification through CHC solving. (Another application to
program synthesis through the notion of oracle-based programming is also discussed
in \iffull Appendix~\ref{sec:oracle}. \fi
 \iffull
We believe that program verification and synthesis are attractive application
domains of neural networks (and machine learning in general), as training data
can be automatically collected.
\fi
%
We plan to extend our NN-guided synthesis tool to enable the synthesis of (i) functions returning
non-Boolean values (such as integers), (ii) predicates/functions on
recursive data structures, and (iii) program expressions containing
loops. For (ii) and (iii), we plan to deploy recurrent neural
networks.


\subsubsection*{Acknowledgments}
We would like to thank anonymous referees for useful comments.
This work was supported by
JSPS KAKENHI Grant Numbers JP20H05703, JP20H04162, and JP19K22842, and
ERATO HASUO Metamathematics for Systems Design Project (No.\ JPMJER1603), JST.

%
%

\appendix
\section*{Appendix}
\section{Additional Information for Section~\ref{sec:PN}}
\label{app:sec2}
\iffull
\subsection{Instances for Learning $(A\land B)\lor (C\land D)$}
\label{sec:2d4instances}
Below is the visualization of all the 20 instances used in the experiments for
learning formulas of the form $(A\land B)\lor (C\land D)$.
The small circles represent positive examples, and the white area is (implicitly) filled with
negative examples.
\begin{minipage}{6cm}
\ \\
  Instance \#1:\\
  $(x-4y-9>0\land 2x-y-9>0)$\\
  $\lor (-x+y-1>0\land x+2y+2>0)$\\

\end{minipage}
\subsection{Instances for Quadratic Constraints}
\label{sec:deg2instances}
\input{data2}

\fi
\subsection{On Three-Layer vs Four-Layer NNs}
\label{sec:twolayer-patchwork}

As reported in Section~\ref{sec:exp},
three-layer NNs with a sufficient number of nodes performed well in the experiment
on learning formulas \((A\land B)\lor (C\land D)\), contrary to our expectation.
This section reports our analysis to find out the reason.

\begin{figure}
\begin{minipage}{6cm}  
  \begin{center}
    \unitlength=0.3mm

  \end{center}
  \end{minipage}
  \begin{minipage}{6cm}
  \begin{center}
    \includegraphics[scale=0.25]{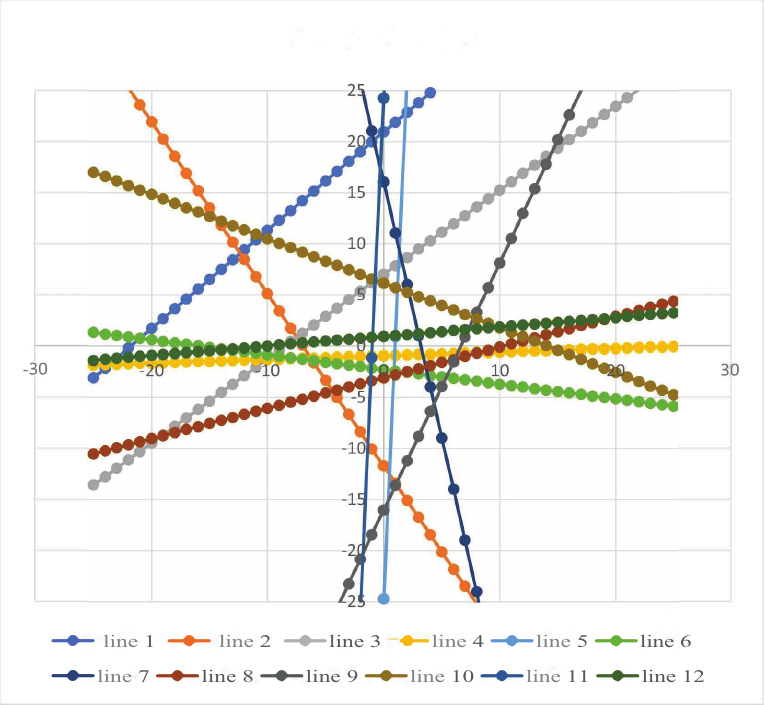}
    \end{center}
  \end{minipage}
  \caption{Problem Instance for Learning $(x\ge 0\land y\ge 0)\lor (x\le 0\land y\le 0)$
  (left) and Lines Recognized by Three-Layer NNs for
  Problem Instance $(x\ge 0\land y\ge 0)\lor (x\le 0\land y\le 0)$ (right).}
\label{fig:test0}
\end{figure}

We used the instance shown on the left-hand side of Figure~\ref{fig:test0},
and compared the training results of three-layer and four-layer NNs.
To make the analysis easier, we tried to train the NNs with a minimal number of hidden nodes
in the second layer.
For the four-layer-case, the training succeeded for only two hidden nodes in the second layer
(plus eight hidden nodes in the third layer), and
only relevant quantifiers of the form \(x>c, x<c, y>c, y<c\) for
\(c\in \set{-2,1,0,1}\) were generated.
In contrast, for the three-layer case, 12 hidden nodes were required for the training to succeed.
The right-hand side of Figure~\ref{fig:test0}
shows the lines \(b_i+w_{i,x}x + w_{i,y}y=0\; (i\in \set{1,\ldots,12})\)
where \(w_{i,x}, w_{i,y}\) and \(b_i\) are the weights and bias for the \(i\)-th hidden node.
We can see that the lines that are (seemingly) irrelevant to the original formula are
recognized by the hidden nodes. Removing any hidden node
(e.g., the node corresponding to line 1) makes the NN fail to properly separate positive
and negative examples. Thus, the three-layer NN is recognizing positive and negative examples in a
manner quite different from the original formula; it is performing a kind of patchwork to
classify the examples. Nevertheless, even in that patchwork, there are lines close to
the horizontal and vertical lines \(y=0\) and \(x=0\). Thus, if we use the trained NN only to extract qualifier candidates
(rather than to recover the whole formula by inspecting also the weights of the third layer),
three-layer NNs can be useful, as already observed in the experiments in Section~\ref{sec:exp}.

  \subsection{On Activation Functions}
\label{sec:act}

To justify our choice of the sigmoid function as activation functions,
we have
replaced the activation functions with
ReLU (\(f(x)=x\) for \(x\ge 0\) and \(f(x)=0\) for \(x<0\)) and
Leaky ReLU (\(f(x)=x\) for \(x\ge 0\) and \(f(x)=-0.01x\)  for \(x<0\))
and conducted the experiments for synthesizing \((A\land B)\lor (C\land D)\)
by using the same problem instances as those used in Section~\ref{sec:exp}.
Here are the results.
\begin{center}
\begin{tabular}{|c|c|c|c|c|c|c|}
\hline   
activation & \#hidden nodes & \#retry & \%success & \%qualifiers & \#candidates
& time (sec.)\\
\hline
\hline
ReLU & 32:32 & 15 & 18.3\% & 66.7\% & 72.4 & 55.9\\
\hline
Leaky ReLU & 32:32 & 32 & 61.7\% & 88.8\% & 64.4 & 52.8\\
\hline
sigmoid & 32:32 & 10 & 85.0\% & 91.7\% & 40.4 & 39.9\\
\hline
\end{tabular}
\end{center}
In this experiment, we have used the mean square error function as the loss function
(since the log loss function assumes that the output belongs to \([0,1]\),
which is not the case here). For comparison, we have re-shown the result
for the sigmoid function (with the mean square error loss function).

As clear from the table above, the sigmoid function
performs significantly better than ReLU and Leaky ReLU, especially
in \%success (the larger is better) and \#candidates (the smaller is better).
This confirms our expectation that the use of the sigmoid function
helps us to ensure that 
only information about
\(b+w_1\,x_1+\cdots+w_k\,x_k>c\) for small \(c\)'s may be propagated
to the output of the second layer, so that we can find suitable
qualifiers by looking at only the weights and biases for the hidden nodes in
the second layer. We do not report experimental results for
the tanh function (\(\texttt{tanh}(x) = 2\sigma(x)-1\)), but
it should be as good as the sigmoid function, as it has similar characteristics.

\subsection{On Biases}
\label{sec:bias}
As for the biases in the second later, we have actually removed them
and instead added a constant input \(1\), so that the weights for
the constant input play the role of the biases (thus, for two-dimensional
input \((x,y)\), we actually gave three-dimensional input \((x,y,1)\)).
This is because, for some qualifier that requires a large constant
(like \(x+y-150>0\)), adding additional constant inputs such as \(100\)
(so that inputs are now of the form \((x,y,1,100)\))
makes the NN training easier to succeed. Among the experiments reported
in this paper, we added an additional constant input \(10\)
for the experiments in Section~\ref{sec:nonlinear}.

Similarly, we also observed (in the experiments not reported here)
that, when the scales of inputs vary extremely among different dimensions
(like \((x,y) = (1, 100), (2, 200), ...\)), then the normalization of
the inputs helped the convergence of training.

  \section{Additional Information for Section~\ref{sec:ICE}}
\label{app:sec3}

Here we provide more details about our experiments on the CHC problem \texttt{xyz},
which shows a general pitfall of the ICE-based CHC solving approach
of \hoice{} (rather than that of our
neural network-guided approach).
Here is the source program of \texttt{xyz} written in OCaml (we have simplified
the original program, by removing redundant arguments).
\begin{verbatim}
let rec loopa x z = 
    if (x < 10) then loopa (x + 1) (z - 2) else z

let rec loopb x z = 
    if (x > 0) then loopb (x-1) (z+2) else assert(z > (-1))

let main (mm:unit(*-:{v:Unit | unit}*)) =
    let x = 0 in let z = 0 in
    let r = loopa x z in
    let s = 10 in loopb s r
\end{verbatim}
Here is (a simplified version of) the corresponding CHC generated by \rtype{}.
\newcommand\loopa{\mathit{loopa}}
\newcommand\loopb{\mathit{loopb}}
\begin{eqnarray}
  x< 10\land \loopa(x+1, z-2, r) &\imp& \loopa(x, z, r) \label{chc:e}\\
  x\ge 10 &\imp& \loopa(x, z, z) \label{chc:d}\\
  x>0 \land \loopb(x,z) &\imp& \loopb(x-1,z+2)  \label{chc:c}\\
  x\le 0 \land \loopb(x,z) &\imp& z>-1  \label{chc:b}\\
  \loopa(0,0,r)&\imp& \loopb(10,r) \label{chc:a}
\end{eqnarray}
When we ran \hoice{} for collecting learning data, 
we observed that \emph{no} positive examples
for \(\loopb\) and \emph{no} negative examples for \(\loopa\) were collected.
Thus \NNG{} returns a trivial solution such as \(\loopa(x,z,r)\equiv \true\) and
\(\loopb(x,z)\equiv \false\).
The reason why \hoice{} generates no positive examples for \(\loopb\) is as follows.
A positive example of \(\loopb\) can only be generated from
the clause (\ref{chc:a}), only when a positive
example of the form \(\loopa(0,0,r)\)
is already available. To generate a positive example of the form \(\loopa(0,0,r)\),
however, one needs to properly instantiate the clauses
(\ref{chc:d}) and (\ref{chc:e}) repeatedly; since \hoice{} generates examples only
lazily when a candidate model returned by the learner does not satisfy
the clauses. In short, \hoice{} must follow a very narrow sequence of
non-deterministic choices to generate the first positive example of \(\loopb\).
Negative examples of \(\loopa\) are rarely generated for the same reason.

Another obstacle is that even if \hoice{} can 
generate a negative counterexample
through clause (\ref{chc:a}) with a luck, it is only of the form \(\loopa(0,0,r)\).
Although further negative examples can be generated through (\ref{chc:e}), the shape
of the resulting negative examples are quite limited.

\iffull
To deal with the problem above, we have prepared an option to collect learning data
also directly from CHCs, by unit propagation.
For example, the negative example \(\loopb(0,-1)\imp \false\) can be generated from
the clause \ref{chc:b} above. By combining it with the clause \ref{chc:c},
negative examples \(\loopb(1,-3)\imp, \ldots, \loopb(10, -21)\imp\) are generated.
By combining \(\loopb(10, -21)\imp\) with the clause \ref{chc:a},
the negative example \(\loopa(0,0,-21)\imp\) can be generated.
As reported in Section~\ref{sec:ICE}, with this option enabled, our tool could
successfully solve \texttt{xyz}.

The variant \texttt{xyz\_v} comes from the following variation of the OCaml program,
obtained by generalizing \(10\) and \(0\) to arbitrary integer values.
Thus, from the viewpoint of program verification, the variant is harder to verify: in fact,
Z3 times-out for this variant. Our tool, however, actually runs faster for the variant, 
because negative examples are easier to collect for this variant.
\begin{verbatim}
let rec loopa b x z = 
    if (x < b) then loopa b (x + 1) (z - 2) else z

let rec loopb x z = 
    if (x > 0) then loopb (x-1) (z+2) else z

let main b z = 
    let x = 0 in 
    let r = loopa b x z in
    let s = b in assert(loopb s r >= z)
\end{verbatim}
\fi

\iffull  
\iffull
\section{Preliminary Experiment on Programming with Oracles}
\label{sec:oracle-exp}
\fi
We have picked several OCaml programs from the
benchmark set in \url{https://github.com/hopv/benchmarks}, and tested our framework.
Some of them are reported below.

Consider the following program.
\begin{verbatim}
let rec sum n =
  if n <= 0 then 0 else n + sum (n-1)
let main n = assert (n <= sum n)
\end{verbatim}
We have replaced the conditional expression with an oracle, and obtained:
\begin{verbatim}
let rec sum n =
  if oracle(n) then 0 else n + sum (n-1)
let main n = assert (n <= sum n)
\end{verbatim}
The goal is now to check whether the original expression for \texttt{oracle(n)}
(i.e., \verb|n<=0|) can be recovered.

To generate training data for \texttt{oracle}, we used random testing.
For that purpose, the code above was surrounded with the following code.
\begin{verbatim}
let history = ref []
let record (i,o) = history := (i,o)::!history
let gen_constraint() = ... 
let oracle(n) = let b = Random.int(2)>0 in (record(n,b); b)

... (* the program above *)

let gen_ex() =  (* testing *)
  for n= -10 to 10 do
    (history := []; 
     try
       main n
     with Assertion_failure _ -> gen_constraint())
  done
\end{verbatim}
The oracle generates a random Boolean \(b\), records it in the call history,
and returns \(b\).
Upon an assertion failure, \verb|gen_constraint()| is called,
which generates implication constraints.
By running \verb|gen_ex()| twice, we obtained the following examples.
\[
\begin{array}{l}
\mbox{ Negative examples:}\\
\quad \oracle(1), \oracle(3),
    \oracle(5),     \oracle(6),     \oracle(7),     \oracle(8),
    \oracle(9),\\\quad     \oracle(10)\\
\mbox{ Implication examples: }\\
\quad \oracle(-8)\imp \oracle(-7)\lor \oracle(-6)\\
\quad\oracle( -12)  \imp \oracle( -11 )\lor \oracle( -10 )\lor \oracle( -9 )\lor \oracle( -8 )\\
\quad\qquad\qquad\qquad\lor \oracle( -7 )\lor \oracle( -6 )\lor \oracle( -5 )\\
\quad \oracle( -6 ) \imp  \oracle( -5 )\lor \oracle( -4 )\lor \oracle( -3 )\lor \oracle( -2 )\\
\quad \oracle( -2 ) \imp  \oracle( -1 )\lor \oracle( 0 )\lor \oracle( 1 )\\
\quad \oracle( -10)  \imp  \oracle( -9 )\lor \oracle( -8 )\lor \oracle( -7 )\\
\quad \oracle( -3 ) \imp  \oracle( -2 )\lor \oracle( -1 )
\end{array}
\]

Given the examples above, our synthesis tool returned \(\oracle(n)\equiv \false\).
(Note that it indeed satisfies all the constraints above.)

Then, the resulting program is:
\begin{verbatim}
let rec sum n =
  if false then 0 else n + sum (n-1)
let main n = assert (n <= sum n)
\end{verbatim}
Obviously, the program does not fail, but never terminates.
By testing the program above, we obtain an infinite sequence of oracle calls, like:
\[
\oracle(0)=\false; \oracle(-1)=\false; \oracle(-2)=\false; \cdots
\]
We have cut the infinite sequence, and added the following implication example:
\[
\imp \oracle(0)\lor \oracle(-1)\lor\oracle(-2)\lor \cdots\lor \oracle(-5).
\]
By running our synthesis tool again, we have obtained
\(\oracle(n) \equiv n<1\).
Thus, we obtained 
\begin{verbatim}
let rec sum n =
  if n<1 then 0 else n + sum (n-1)
let main n = assert (n <= sum n)
\end{verbatim}
and the resulting program could be verified by MoCHi (\url{https://github.com/hopv/MoCHi}), an automated program verification tool for functional programs.

Let us consider another example.
The following program has also been 
taken from the same benchmark set, where the original 
conditional expression \(x\ge y\) in \verb|f| has been replaced by
an oracle call.
we have also replaced the specification as the original assertion
\verb|f x m = m| admits a silly solution \(\oracle(x,y)=\false\).
\begin{verbatim}
let max max2 (x:int) (y:int) (z:int) : int = max2 (max2 x y) z

let f x y : int = if oracle(x,y) then x else y

let main (x:int) y z =
  let m = max f x y z in
  assert (m>=x && m >= y && m>=z)
\end{verbatim}
As before, the above code was surrounded with the code for testing the code
and generating implication examples.
After running the test several times,
the following implication examples were accumulated.
\[
\begin{array}{l}
\oracle(1, 3)\land \oracle(1, 7) \imp\\
\oracle(2, -5)\imp  \oracle(2, -10)\\
\imp \oracle(8, -4)\lor \oracle(-3, 8)\\
\oracle(6, 1)\imp \oracle(6, -4)\\
\imp \oracle(-10, -4)\lor  \oracle(3, -10)\\
\cdots
\end{array}
\]
Given those examples, our synthesis tool generated
\(\oracle(x,y) \equiv 2x-3y-1>0\), which is a wrong solution.
By testing the program obtained by inserting the wrong oracle:
\begin{verbatim}
let max max2 (x:int) (y:int) (z:int) : int = max2 (max2 x y) z

let f x y : int = if 2*x + 3*y -1 >0 then x else y

let main (x:int) y z =
  let m = max f x y z in
  assert (m>=x && m >= y && m>=z)
\end{verbatim}
we obtained further implication examples:
\[
\begin{array}{l}
\oracle(9, -6)\imp \oracle(9, 8)\\
\oracle(8, 0)\imp \oracle(8, 6).
\end{array}
\]
After adding them to the training data, our tool generated
\(\oracle(x,y) \equiv x-y+1>0\), which is correct.
The resulting program could be verified by MoCHi.

As a little more non-trivial example, we have also tested McCarthy's 91 function:
\begin{verbatim}
let rec mc91 x = if oracle(x) then x - 10 else mc91 (mc91 (x + 11))
let main n = if n <= 101 then assert (mc91 n = 91)
\end{verbatim}
As before, we have first replaced \verb|oracle(x)| with a non-deterministic Boolean function,
and collected training data by randomly executing the program.
From runs that lead to assertion failures, we could obtain
negative examples of the form \(\oracle(n)\imp\) for \(0\le n\le 100\) (besides
other implication examples).
In addition, we have collected positive examples from successful runs that contain
only a \emph{single} oracle call. Actually,
the only positive example collected was \(\oracle(101)\).
From those data, our tool could correctly synthesize the formula \(x>100\).

\fi
\end{document}